\input{epsf}
\documentstyle[12pt]{article}

\def\simge{\mathrel{%
   \rlap{\raise 0.511ex \hbox{$>$}}{\lower 0.511ex \hbox{$\sim$}}}}
\def\simle{\mathrel{
   \rlap{\raise 0.511ex \hbox{$<$}}{\lower 0.511ex \hbox{$\sim$}}}}
 
\def\slashchar#1{\setbox0=\hbox{$#1$}           
   \dimen0=\wd0                                 
   \setbox1=\hbox{/} \dimen1=\wd1               
   \ifdim\dimen0>\dimen1                        
      \rlap{\hbox to \dimen0{\hfil/\hfil}}      
      #1                                        
   \else                                        
      \rlap{\hbox to \dimen1{\hfil$#1$\hfil}}   
      /                                         
   \fi}                                         %
\def\nn{\nonumber}
\def\ts{\thinspace}

\def\ra{\rightarrow}

\def\ol{\bar}

\def\be{\begin{equation}} 
\def\ee{\end{equation}} 
\def\bea{\begin{eqnarray}}
\def\eea{\end{eqnarray}}
\def\ba{\begin{array}}
\def\ea{\end{array}}
\def\chipr{\chi^{\ts \prime}}
\def\CA{{\cal A}}
\def\CB{{\cal B}}
\def\CC{{\cal C}}
\def\CD{{\cal D}}

\def\CF{{\cal F}}
\def\CG{{\cal G}}

\def\CM{{\cal M}}

\def\CR{{\cal R}}

\def\etmiss{\slashchar{E}_T}

\def\ecm{\sqrt{s}}
\def\shat{\hat s}
\def\that{\hat t}
\def\uhat{\hat u}
\def\rshat{\sqrt{\shat}}

\def\atro{\alpha_{\tro}}

\def\Ntc{N_{TC}}

\def\thw{\theta_W}
\def\kslash{\raise.15ex\hbox{/}\kern-.57em k}
\def\tro{\rho_{T}}

\def\tropm{\rho_{T}^\pm}
\def\trop{\rho_{T}^+}
\def\trom{\rho_{T}^-}
\def\troz{\rho_{T}^0}
\def\tom{\omega_T}
\def\tpi{\pi_T}
\def\tpipm{\pi_T^\pm}
\def\tpimp{\pi_T^\mp}
\def\tpip{\pi_T^+}
\def\tpim{\pi_T^-}
\def\tpiz{\pi_T^0}
\def\tpipr{\pi_T^{0 \prime}}

\def\mev{{\rm MeV}}
\def\gev{{\rm GeV}}
\def\tev{{\rm TeV}}

\def\pb{{\rm pb}}

\def\ifb{{\rm fb}^{-1}}
\def\half{{\textstyle{ { 1\over { 2 } }}}}

\def\fourth{{\textstyle{ { 1\over { 4 } }}}}

\begin{document}
\title{
\vskip -15mm
\begin{flushright}
\vskip -15mm
{\small BUHEP-99-4\\
hep-ph/9903369\\}
\vskip 5mm
\end{flushright}
{\Large{\bf Technihadron Production and Decay in Low-Scale Technicolor}}\\
}
\author{
{\large Kenneth Lane\thanks{lane@buphyc.bu.edu}}\\
{\large Department of Physics, Boston University}\\
{\large 590 Commonwealth Avenue, Boston, MA 02215}\\
}
\maketitle
\begin{abstract}
We discuss the production and decay rates of the lightest color-singlet
technihadrons, spin-one $\tro$ and $\tom$ and spin-zero $\tpi$, in a minimal
``straw-man'' model of low-scale techicolor.  The revised $\tro$ and $\tom$
decay rates affect the technicolor searches planned for Run~II of the
Tevatron Collider.
\end{abstract}


\newpage

\section{Introduction}

Modern technicolor models of dynamical electroweak symmetry breaking require
a walking technicolor gauge coupling~\cite{wtc} to evade large
flavor-changing neutral current effect and the assistance of topcolor
interactions that are strong near 1~TeV~\cite{topcondref,tctwohill,tctwoklee}
to provide the large mass of the top quark. Both additions to the basic
technicolor scenario~\cite{tc,etc} tend to require a large number $N_D$ of
technifermion doublets. Many technifermions are needed to make the beta
function of walking technicolor small. And many seem to be required in
topcolor-assisted technicolor to generate the hard masses of quarks and
leptons, to induce the correct mixing between heavy and light quarks, and to
break topcolor symmetry down to ordinary color. As has been
emphasized~\cite{multi,elw}, large $N_D$ implies a relatively low
technihadron mass scale, set by the technipion decay constant $F_T \simeq
F_\pi/\sqrt{N_D}$, where $F_\pi = 246\,\gev$. In the models of
Ref.~\cite{tctwoklee}, for example, $N_D \simeq 10$ and $F_T \simeq
80\,\gev$. It is likely that this low-scale technicolor will be within reach
of the Tevatron Collider Run~II experiments.~\footnote{The Run~II conditions
assumed in this paper are $p \ol p$ collisions at center-of-mass energy
$\sqrt{s} = 2\,\tev$ for an integrated luminosity of $2\,\ifb$.} Indeed,
preliminary searches based on Run~I data have been carried out or are in
progress for several of its color-singlet signals~\cite{searcha, searchb,
searchc}.

In this paper we re-examine the decay and production rates for color-singlet
technivector mesons, $V_T = \tro$ and $\tom$. Special attention is given to
the decay $V_T \ra G \tpi$, where $G$ is a transversely polarized electroweak
gauge boson, $\gamma$, $Z^0$, $W^\pm$, and $\tpi$ is a technipion. The gauge
boson polarization is defined relative to the spin direction of the
technivector meson in the latter's rest frame. (This is the same as the beam
direction in a hadron or lepton collider.) Some of these decay rates,
particularly those involving a photon, can be as large as the modes
previously considered~\cite{elw}. If this happens, branching ratio
expectations are different from Ref.~\cite{elw} and the limits placed by
analyses in Ref.~\cite{searcha,searchb,searchc} must be reinterpreted. In any
case, signal rates are large enough that technicolor searches in Run~II will
severely restrict the expected parameter space of low-scale technicolor.

To set the ground rules for our calculations, we adopt the ``Technicolor
Straw Man Model''. In the TCSM, we assume that we can consider in isolation
the lowest-lying bound states of the lightest technifermion doublet, $(T_U,
T_D)$. These technifermions are likely to be color singlets because
color-$SU(3)$ interactions contribute significantly to their hard
mass~\cite{multi}. We shall assume that they transform under technicolor
$SU(\Ntc)$ as fundamentals. Their electric charges are $Q_U$ and $Q_D$. The
bound states in question are vector and pseudoscalar mesons. The vectors
include a spin-one isotriplet $\tro^{\pm,0}$ and an isosinglet $\tom$. In
topcolor-assisted technicolor, there is no need to invoke large
isospin-violating extended technicolor interactions to explain the top-bottom
splitting. Thus, techni-isospin can be, and likely must be, a good
approximate symmetry. Then, $\tro$ and $\tom$ will be nearly
degenerate. Their mixing will be described in the neutral-sector propagator
matrix, $\Delta_0$, in Eq.~(\ref{eq:gzprop}) below.

The lightest pseudoscalar $(T_U,T_D)$ bound states, the technipions, also
comprise an isotriplet $\Pi_T^{\pm,0}$ and an isosinglet $\Pi_T^{0
\prime}$. However, these are not mass eigenstates. In the TCSM, we assume the
isovectors are simple two-state mixtures of the longitudinal weak bosons
$W_L^\pm$, $Z_L^0$---the true Goldstone bosons of dynamical electroweak
symmetry breaking in the limit that the $SU(2) \otimes U(1)$ couplings $g,g'$
vanish---and mass-eigenstate pseudo-Goldstone technipions $\tpi^\pm, \tpiz$:
\be\label{eq:pistates}
 \vert\Pi_T\rangle = \sin\chi \ts \vert
W_L\rangle + \cos\chi \ts \vert\tpi\rangle\ts.
\ee
Here, $\sin\chi = F_T/F_\pi \ll 1$.

Similarly, $\vert\Pi_T^{0 \prime} \rangle = \cos\chipr \ts
\vert\tpipr\rangle\ + \cdots$, where $\chipr$ is another mixing angle and the
ellipsis refer to other technipions needed to eliminate the two-technigluon
anomaly from the $\Pi_T^{0 \prime}$ chiral current. It is unclear whether,
like $\tro$ and $\tom$, these neutral technipions will be degenerate as we
have previously supposed~\cite{elw}. On one hand, they both contain the
lightest $\ol T T$ as constituents. On the other, $\tpipr$ must contain
other, presumably heavier, technifermions as a consequence of anomaly
cancellation. In our calculations, we shall assume as before that $\tpiz$ and
$\tpipr$ are nearly degenerate. We reiterate the point made in
Ref.~\cite{elw} that, if they are and if their widths are roughly equal,
there will be appreciable $\tpiz$--$\tpipr$ mixing. Then, the lightest
neutral technipions will be ideally-mixed $\ol T_U T_U$ and $\ol T_D T_D$
bound states. In any case, the technipions are expected to decay as follows:
$\tpip \ra c \ol b$ or $c \ol s$ or even $\tau^+ \nu_\tau$; $\tpiz \ra b \ol
b$ and, perhaps $c \ol c$, $\tau^+\tau^-$; and $\tpipr \ra gg$, $b \ol b$, $c
\ol c$, $\tau^+\tau^-$.~\footnote{All technihadron decay and production rates
in the TCSM are compiled for easy reference in a companion to this
paper, Ref.~\cite{tcsmrates}.}

In the limit that the electroweak couplings $g,g' = 0$, the $\tro$ and $\tom$
decay as
\bea\label{eq:vt_decays}
\tro &\ra& \Pi_T \Pi_T = \cos^2 \chi\ts (\tpi\tpi) + 2\sin\chi\ts\cos\chi
\ts (W_L\tpi) + \sin^2 \chi \ts (W_L W_L) \ts; \nn \\\nn\\
\tom &\ra& \Pi_T \Pi_T \Pi_T = \cos^3 \chi \ts (\tpi\tpi\tpi) + \cdots \ts.
\eea
The $\tro$ decay amplitude is
\be\label{eq:rhopipi}
\CM(\tro(q) \ra \pi_A(p_1) \pi_B(p_2)) = g_{\tro} \ts \CC_{AB}
\ts \epsilon(q)\cdot(p_1 - p_2) \ts,
\ee
where $\epsilon(q)$ is the $\tro$ polarization vector; $\atro \equiv
g_{\tro}^2/4\pi = 2.91(3/\Ntc)$ is scaled naively from QCD and $\Ntc = 4$ is
used in calculations; and
\be\label{eq:ccab}
\ba{ll}
\CC_{AB} &= \left\{\ba{ll} \sin^2\chi  & {\rm for} \ts\ts\ts\ts W_L^+ W_L^-
\ts\ts\ts\ts {\rm or} \ts\ts\ts\ts  W_L^\pm Z_L^0 \\
\sin\chi \cos\chi & {\rm for} \ts\ts\ts\ts W_L^+ \tpim, W_L^- \tpip
\ts\ts\ts\ts  {\rm or} \ts\ts\ts\ts W_L^\pm \tpiz, Z_L^0 \tpipm \\
\cos^2\chi & {\rm for} \ts\ts\ts\ts \tpip\tpim  \ts\ts\ts\ts {\rm or}
\ts\ts \ts\ts \tpipm\tpiz \ts.
\ea \right.
\ea
\ee
The $\tro$ decay rate to two technipions is then (for later use in cross
sections, we quote the energy-dependent width for a $\tro$ mass of
$\sqrt{\shat}$)
\be\label{eq:trhopipi}
\Gamma(\troz \ra \pi_A^+ \pi_B^-) = \Gamma(\tropm \ra \pi_A^\pm \pi_B^0) =
{2 \atro \CC^2_{AB}\over{3}} \ts {\ts\ts p^3\over {\shat}} \ts,
\ee
where $p = [(\shat - (M_A+M_B)^2) (\shat - (M_A-M_B)^2)]^\half/2\rshat$
is the $\tpi$ momentum in the $\tro$ rest frame.

Now, walking technicolor enhancements of technipion masses are likely to
close off the channels $\tro \ra \tpi\tpi$, $\tom \ra \tpi\tpi\tpi$ and even
the isospin-violating $\tom \ra \tpi\tpi$~\cite{multi}. A technirho of, say,
200~GeV may then decay to $W_L \tpi$ or $W_L W_L$, but how does a light
techniomega decay? The answer is that all its decays are electroweak, $\tom
\ra \gamma\tpiz$, $Z^0\tpiz$, $W^\pm \tpimp$, etc., where $Z$ and $W$ may be
either transversely or longitudinally polarized. This raises the further
question: Since $\sin^2\chi \ll 1$, the electroweak decays of $\tro$ to the
transverse gauge bosons $\gamma,W,Z$ plus a technipion may be competitive
with the open-channel strong decays. How do we correctly describe these $g,g'
\neq 0$ transitions?  If the rates for these radiative decays are not
negligible, they affect expectations for the existing and planned
searches for $\tro \ra W\tpi$, $\tom \ra \gamma\tpiz$ and $\tro,\tom \ra
f_i\ol f_i$.

In Section~2, we discuss the form of the amplitudes for the decays $V_T \ra
G\tpi$ where $G=\gamma,Z^0, W^\pm$ is transversely polarized. We shall see
that, depending on the size of technicolor-scale mass parameters $M_{V,A}$
and technifermion charges $Q_{U,D}$, several of these decays have rates as
large as those considered in Ref.~\cite{elw}. In Section~3, we present the
cross sections for all $q \ol q \ra \tro, \tom \ra X$ subprocesses of
interest in the color-singlet sector of the TCSM. Section~4 contains a sample
of numerical results for $\tro$ and $\tom$ signal rates in $p \ol p$
collisions at $\ecm = 2\,\tev$.

\section{$\tro, \tom \ra \gamma/W/Z + \tpi$ when $g,g'\neq 0$}

It is simplest to start with the decay $\tom \ra \gamma\tpiz$ considered
already in Ref.~\cite{elw}. Gauge invariance, chiral symmetry, angular
momentum and parity conservation imply that the lowest-dimensional operator
mediating this decay is $(e/M_V)\ts F_{\tro} \cdot \widetilde{F}_\gamma \ts
\tpiz$ where, naively scaling from analogous decays in QCD, $M_V$ is a
parameter of order several 100~GeV.~\footnote{The corresponding $\rho \ra
\gamma\pi$ parameter in QCD is about $400\,\mev$. A large--$N_c$ argument
implies $M_V \simeq (F_T/f_\pi) \ts 400\,\mev \simeq 350\,\gev$.} To fix its
normalization, we write this decay amplitude as
\be\label{eq:omgampi}
\CM(\tom(q) \ra \gamma(p_1) \tpiz(p_2)) = {e \cos\chi\over{M_V}}
\epsilon^{\mu\nu\lambda\rho} \epsilon_\mu(q) \epsilon^*_\nu(p_1) q_\lambda
p_{1\rho} \ts.
\ee

It is now clear on dynamical and symmetry grounds that the amplitude for
decay to any transversely polarized electroweak boson $G$ plus a technipion
can be written as
\bea\label{eq:genamp}
\CM(V_T(q) \ra G(p_1) \tpi(p_2)) &=& {eV_{V_T G\tpi} \over{M_V}}\ts
\epsilon^{\mu\nu\lambda\rho} \ts \epsilon_\mu(q) \ts \epsilon^*_\nu(p_1) \ts
q_{\lambda} \ts p_{1\rho} \\
 &+& {eA_{V_T G\tpi} \over{M_A}} \biggl(\epsilon(q) \cdot \epsilon^*(p_1)\ts
q\cdot p_1 - \epsilon(q) \cdot p_1\ts \epsilon^*(p_1) \cdot q\biggr) \nn
\ts.
\eea
The first term corresponds to the vector coupling of $G$ to the constituent
technifermions of $V_T$ and $\tpi$ and the second term to its axial-vector
coupling. Note that the amplitudes for emission of longitudinally polarized
bosons in Eq.~(\ref{eq:rhopipi}) and transversely polarized ones in
Eq.~(\ref{eq:genamp}) are noninterfering, as they should be. On dynamical
grounds, the mass parameter $M_A$ is expected to be comparable to $M_V$. If
we adopt a ``valence technifermion'' model for the graphs describing
Eq.~(\ref{eq:genamp})---a model which works very well for $\omega, \rho \ra
\gamma \pi$ and $\gamma \eta$ in QCD---CP-invariance implies that the $V$ and
$A$ coefficients in this amplitude are given in our normalization
by~\footnote{We have neglected decays such as $\troz \ra W_T W_L$ and $\troz
\ra W_T W_T$. The rate for the former is suppressed by $\tan^2 \chi$ relative
to the rate for $\troz \ra W_T \tpi$ while the latter's rate is suppressed by
$\alpha$.}
\be\label{eq:VA}
V_{V_T G\tpi} = {\rm Tr}\biggl(Q_{V_T} \{Q^\dagger_{G_V}, \ts
Q^\dagger_{\tpi}\}\biggr) \ts,\qquad
A_{V_T G\tpi} = {\rm Tr}\biggl(Q_{V_T} [Q^\dagger_{G_A}, \ts
Q^\dagger_{\tpi}]\biggr) \ts.
\ee
In the TCSM, with electric charges $Q_U$, $Q_D$ for $T_U$, $T_D$, the
generators $Q$ in Eq.~(\ref{eq:VA}) are given by
\bea\label{eq:charges}
Q_{\troz} &=& {1\over{\sqrt{2}}} \left(\ba{cc} 1 & 0 \\ 0 & -1 \ea\right)
\ts;\qquad
Q_{\trop} = Q^\dagger_{\trom} = \left(\ba{cc} 0 & 1 \\ 0 & 0 \ea\right)\nn\\
Q_{\tpiz} &=& {\cos\chi\over{\sqrt{2}}} \left(\ba{cc} 1 & 0 \\ 0 & -1 \ea\right)
\ts;\ts\quad
Q_{\tpip} = Q^\dagger_{\tpim} = \cos\chi\left(\ba{cc} 0 & 1 \\ 0 & 0
  \ea\right)\nn\\
Q_{\tpipr} &=& {\cos\chipr\over{\sqrt{2}}} \left(\ba{cc} 1 & 0 \\ 0
  & 1 \ea\right) \nn\\ \nn\\
%
Q_{\gamma_V} &=& \left(\ba{cc} Q_U& 0 \\ 0 & Q_D \ea\right)
\ts;\qquad
Q_{\gamma_A} = 0 \nn\\
Q_{Z_V} &=& {1\over{\sin\thw \cos\thw}} \left(\ba{cc} \fourth - Q_U
  \sin^2\thw & 0 \\ 0 & -\fourth - Q_D \sin^2\thw \ea\right) \nn\\
Q_{Z_A} &=& {1\over{\sin\thw \cos\thw}} \left(\ba{cc} -\fourth & 0 \\ 0
  &  \fourth \ea\right) \nn\\
Q_{W^+_V} &=& Q^\dagger_{W^-_V} = -Q_{W^+_A} = -Q^\dagger_{W^-_A} = {1\over
  {2\sqrt{2}\sin\thw}}\left(\ba{cc} 0 & 1 \\ 0 & 0 \ea\right)
\eea

The decay rate for $V_T \ra G \tpi$ is
\be
\Gamma(V_T \ra G \tpi) = {\alpha V^2_{V_T G\tpi} \ts p^3\over {3M_V^2}} +
{\alpha A^2_{V_T G\tpi} \ts p\ts(3 M_G^2 + 2p^2)\over {6M_A^2}} \ts,
\ee
where $p$ is the $G$-momentum in the $V_T$ rest frame.
The $V$ and $A$ coefficients and sample decay rates are listed in
Table~1. These are to be compared with the rates for decay into longitudinal
$W$ and $Z$ bosons plus a technipion quoted in Ref.~\cite{elw}. For $M_{\tro}
= 210\,\gev$, $M_{\tpi} = 110\,\gev$, and $\Ntc = 4$, they are
\bea\label{eq:longdecay}
\Gamma(\troz \ra W_L^\pm \tpimp) &=& \Gamma(\tropm \ra W_L^\pm \tpiz) =
2.78\sin^2\chi\cos^2\chi \nn \\
\Gamma(\tropm \ra Z_L^0 \tpimp) &=& 0.89\sin^2\chi\cos^2\chi \ts.
\eea
For $\sin^2\chi = 1/9$, our nominal choice, and for $M_V = M_A = 100\,\gev$,
the rates for $\tro$ and $\tom \ra \gamma\tpi$ and for $\tro \ra W_T \tpi$,
$Z_T\tpi$ via axial vector coupling are comparable to these. Obviously, these
transverse-boson decay rates fall quickly for greater $M_V$ and $M_A$.

We can estimate the rate for the isospin-violating decay $\tom \ra W_T^\pm
\tpimp$ as
\be\label{eq:isov}
\Gamma(\tom \ra W_L^\pm \tpimp) = \vert\epsilon_{\rho\omega}\vert^2 \ts
\Gamma(\troz \ra W_L^\pm \tpimp) \ts,
\ee
where $\epsilon_{\rho\omega}$ is the $\tro$-$\tom$ mixing amplitude. In QCD,
$\vert \epsilon_{\rho\omega}\vert \simeq 5\%$, so we expect this decay mode
to be entirely negligible.

Finally, for completeness, we record here the decay rates for $\tro, \tom \ra
f \ol f$. The $\tro$ decay rates to fermions with $N_f=1$ or 3~colors
are~\footnote{Eqs.~(\ref{eq:trhoff}), (\ref{eq:afactors}) and
(\ref{eq:tomff}) below correct Eqs.~(3) and (6) in the second paper and
Eqs.~(3) and (5) in the third paper of Ref.~\cite{elw}. A factor of
$M^4_{V_T}/\shat^2$ that appears in Eqs.~(6) and~(11) of that second paper
has been eliminated from Eqs.~(\ref{eq:trhoff}) and~(\ref{eq:tomff}). This
convention is consistent with the off-diagonal $s f_{G V_T}$ terms in the
propagator matrices $\Delta_{0,\pm}$ defined in Eqs.~(\ref{eq:gzprop})
and~(\ref{eq:wprop}) below. For weakly-coupled narrow resonances such as
$\tro$ and $\tom$, the difference is numerically insignificant.}
\bea\label{eq:trhoff}
\Gamma(\troz \ra f_i \ol f_i) &=& {N_f \ts \alpha^2 p\over
{3 \atro \shat}} \ts \left((\shat - m_i^2)
\ts A_i^0(\shat) + 6 m_i^2\ts \CR e(\CA_{iL}(\shat)
\CA_{iR}^*(\shat))\right) \ts, \nn\\ \\
\Gamma(\trop \ra f_i \ol f'_i) &=& {N_f \ts \alpha^2 p\over
{6 \atro} \shat^2} \ts \left(2\shat^2 - \shat (m_i^2 + m^{'2}_i) -
(m_i^2 - m^{'2}_i)^2\right) A_i^+(\shat) \ts,\nn
\eea
where a unit CKM matrix is assumed in the second equality. The quantities
$A_i$ are given by
\bea\label{eq:afactors}
A_i^\pm(\shat) &=& {1 \over {8\sin^4\thw}} \ts \biggl\vert{\shat \over {\shat
    - \CM^2_W}}\biggr\vert^2 \ts, \nn \\ \nn\\
A_i^0(\shat) &=& \vert \CA_{iL}(\shat) \vert^2
+ \vert \CA_{iR}(\shat) \vert^2 \ts,
\eea
where, for $\lambda = L,R$,
\bea\label{eq:zfactors}
\CA_{i\lambda}(\shat) &=& Q_i + {2 \zeta_{i\lambda} \ts \cot 2\thw \over
   {\sin 2\thw}} \ts \biggl({\shat \over {\shat - \CM^2_Z}}\biggr)\ts, \nn\\
\zeta_{i L} &=& T_{3i} - Q_i \sin^2\thw \ts, \nn\\
\zeta_{i R} &=& - Q_i \sin^2\thw \ts.
\eea
Here, $Q_i$ and $T_{3i} = \pm 1/2$ are the electric charge and left-handed
weak isospin of fermion $f_i$. Also, $\CM^2_{W,Z} = M^2_{W,Z} - i \rshat \ts
\Gamma_{W,Z}(\shat)$, where $\Gamma_{W,Z}(\shat)$ is the weak boson's
energy-dependent width.~\footnote{Note, for example, that $\Gamma_Z(\shat)$
includes a $t\ol t$ contribution when $\shat > 4m_t^2$.}.

The $\tom$ decay rates to fermions with $N_f$ colors are given by
\be\label{eq:tomff}
\Gamma(\omega_T \ra \ol f_i f_i) = {N_f \ts \alpha^2 p\over
{3 \atro \shat}} \ts \left((\shat - m_i^2)
\ts B_i^0(\shat) + 6 m_i^2\ts \CR e(\CB_{iL}(\shat)
\CB_{iR}^*(\shat))\right) \ts, 
\ee
where
\bea\label{eq:bfactors}
&B_i^0(\shat) &= \vert \CB_{iL}(\shat) \vert^2
+ \vert \CB_{iR}(\shat) \vert^2 \ts, \nn \\ \nn\\
&\CB_{i\lambda}(\shat) &= \left[Q_i - {4 \zeta_{i\lambda} \sin^2\thw \over
    {\sin^2 2\thw}} \biggl({\shat \over {\shat - \CM_Z^2}}\biggr)\right] \ts
(Q_U + Q_D)
\ts.
\eea

{\flushleft{\begin{table}{
\begin{tabular}{|c|c|c|c|c|}
\hline
Process & 
$V_{V_T G\tpi}$ & 
$A_{V_T G\tpi}$ & 
$\Gamma(V_T \ra G_V\tpi)$ & 
$\Gamma(V_T \ra G_A\tpi)$ 
\\
\hline\hline
$\tom \ra \gamma \tpiz$& $c_\chi$ & 0 & 0.115 $c^2_\chi$ & 0  \\
$\ts\ts\ts\quad \ra \gamma \tpipr$ & $(Q_U + Q_D)\ts c_{\chipr}$ & 0 & 0.320
$c^2_{\chipr}$ & 0 \\ 
$\qquad \ra Z^0 \tpiz$ & $c_\chi\cot 2\thw$ & 0 &
$2.9\times 10^{-3}c^2_\chi$ & 0 \\ 
$\ts\qquad \ra Z^0 \tpipr$ & $-(Q_U+Q_D)\ts c_{\chipr}\tan \thw$ & 0 &
$5.9\times 10^{-3}c^2_{\chipr}$ & 0 \\ 
$\ts\ts\ts\ts\qquad \ra W^\pm \tpimp$ & $c_\chi/(2\sin\thw)$ & 0 &
  $2.4\times 10^{-2}c^2_\chi$ & 0 \\ 
\hline
$\troz \ra \gamma \tpiz$ & $(Q_U + Q_D)\ts c_\chi$ & 0 & 0.320 $c^2_\chi$ & 0
\\
$\ts\ts\ts\quad \ra \gamma \tpipr$ & $c_{\chipr}$ & 0 & 0.115 $c^2_{\chipr}$ &
0 \\ 
$\qquad \ra Z^0 \tpiz$ & $-(Q_U+Q_D)\ts c_\chi \tan \thw$ & 0 &
$5.9\times 10^{-3}c^2_\chi$ & 0 \\ 
$\ts\qquad \ra Z^0 \tpipr$ & $c_{\chipr}\ts \cot 2\thw$ & 0 &
$2.9 \times 10^{-3}c^2_{\chipr}$ & 0 \\ 
$\ts\ts\ts\ts\qquad \ra W^\pm \tpimp$ & 0 & $-c_\chi/(2\sin\thw)$ &
  0 & 0.143 $c^2_\chi$  \\ 
\hline
$\tropm \ra \gamma \tpipm$ & $(Q_U + Q_D)\ts c_\chi$ & 0 & 0.320 $c^2_\chi$
& 0 \\ 
$\qquad \ra Z^0 \tpipm$ & $-(Q_U+Q_D)\ts c_\chi \tan \thw$ & $c_\chi
\ts /\sin 2\thw$ & $5.9\times 10^{-3}c^2_\chi$ & 0.147 $c^2_\chi$ \\  
$\ts\ts\ts\qquad \ra W^\pm \tpiz$ & 0 & $c_\chi/(2\sin\thw)$ &
  0 & 0.143 $c^2_\chi$  \\ 
$\ts\ts\ts\qquad \ra W^\pm \tpipr$ & $c_{\chipr}/(2\sin\thw)$ & 0&
  $2.4\times 10^{-2}c^2_{\chipr}$ & 0 \\ 
\hline\hline
\end{tabular}}
\caption{Amplitudes and sample decay rates (in GeV) for $V_T \ra G \tpi$. 
In the rate calculations, $M_{V_T} = 210\,\gev$, $M_{\tpi} = 110\,\gev$, $M_V
= M_A = 100\,\gev$; technifermion charges are $Q_U + Q_D
=\textstyle{5\over{3}}$; $c_\chi = \cos\chi$ and $c_{\chipr} = \cos\chipr$;
$G_V$ and $G_A$ refer to decays involving the vector and axial-vector
couplings, respectively.}
\end{table}}}

\section{Cross Sections}

In this section we record cross sections for the hadron collider subprocesses
$q \ol q \ra V_T \ra \tpi\tpi$, $G\tpi$, and $f \ol f$. All of these these
may be influenced by the fact that the $\tro \ra \gamma\tpi$ decay rates are
comparable to the previously considered $\tom \ra \gamma\tpiz$. Thus, for
example, so long as $\tro$ and $\tom$ are nearly degenerate and the
technipions in question decay to at least one $b$-quark, these additional
modes contribute to the signal of a photon plus dijets with a single $b$-tag
studied in one recent CDF analysis~\cite{searchb}.

As we'll see in the sample calculations in Section~4, it is important to
include $\tro$-$\tom$ interference in these cross sections (also see the
third paper in Ref.~\cite{elw}). In the TCSM, the
$\gamma$--$Z^0$--$\troz$--$\tom$ propagator matrix $\Delta_0$ is the inverse
of
\be\label{eq:gzprop}
\Delta_0^{-1}(s) =\left(\ba{cccc}
s & 0 & -s f_{\gamma\tro} & -s f_{\gamma\tom} \\
0 & s - \CM^2_Z  & -s f_{Z\tro} & -s f_{Z\tom} \\
-s f_{\gamma\tro}  & -s f_{Z\tro}  & s - \CM^2_{\troz} & 0 \\
-s f_{\gamma\tom}  & -s f_{Z\tom}  & 0 & s - \CM^2_{\tom} 
\ea\right) \ts.
\ee
Note that, in the spirit of vector-meson dominance, we are assuming only
kinetic mixing between the gauge bosons and technivector mesons. As noted
earlier, whether this should be $s f_{GV_T}$ or $M^2_{V_T} f_{GV_T}$ is
numerically irrelevant for narrow resonances. In setting the off-diagonal
$\troz$--$\tom$ elemements of this matrix equal zero, we are guided by the
smallness of this mixing in QCD and by the desire to keep the number of
adjustable parameters in the TCSM as small as possible. Of course, such
mixing can always be added if warranted. The properly normalized $G V_T$
couplings are
\be\label{eq:fgv}
f_{GV_T} = \sqrt{2}\ts\xi\ts {\rm Tr}\biggl(Q_{G_V}Q^\dagger_{V_T}\biggr)
\ts;
\ee
in particular, $f_{\gamma\tro} = \xi$, $f_{\gamma\tom} = \xi \ts (Q_U +
Q_D)$, $f_{Z\tro} = \xi \ts \cot 2\thw$, and $f_{Z\tom} = - \xi \ts (Q_U +
Q_D) \tan\thw$, where $\xi = \sqrt{\alpha/\atro}$.
In the charged sector, the $W^\pm$--$\tropm$ matrix is the inverse of
\be\label{eq:wprop}
\Delta_{\pm}^{-1}(s) =\left(\ba{cc} s - \CM^2_W & -s f_{W\tro} \\ -s
  f_{W\tro} & s - \CM^2_{\tropm} \ea\right) \ts,
\ee
where $f_{W\tro} = \xi/(2\sin\thw)$.

The rates for production of any technipion pair, $\pi_A\pi_B = W_L W_L$,
$W_L\tpi$, and $\tpi\tpi$, in the isovector ($\tro$) channel are:
\bea\label{eq:pippim}
& &{d\hat\sigma(q_i \ol q_i \ra \troz \ra \pi^+_A\pi^-_B)
  \over{d\that}} = \nn\\
& & \qquad{\pi \alpha\atro \CC^2_{AB} 
  (4\shat p^2 -(\that-\uhat)^2) \over{12 \shat^2}} \ts
\biggl(\vert\CF^{\tro}_{iL}(\shat)\vert^2 +
  \vert\CF^{\tro}_{iR}(\shat)\vert^2\biggr)\ts,
\eea
and
\be\label{eq:pippiz}
{d\hat\sigma(u_i \ol d_i \ra \trop \ra \pi^+_A\pi^0_B)
  \over{d\that}} = 
{\pi \alpha\atro \CC^2_{AB} 
  (4\shat p^2 -(\that-\uhat)^2) \over{24 \sin^2\thw \shat^2}} \ts
\vert\Delta_{W\tro}(\shat)\vert^2 \ts,
\ee
where $p = [(\shat - (M_A+M_B)^2) (\shat - (M_A-M_B)^2)]^\half/2\rshat$ is the
$\shat$-dependendent momentum of $\pi_{A,B}$. As usual, $\that = M^2_A -
\rshat(E_A - p\cos\theta)$, $\uhat = M^2_A - \rshat(E_A + p\cos\theta)$,
where $\theta$ is the c.m. production angle of $\pi_A$.  The factor $4\shat
p^2 -(\that-\uhat)^2 = 4 \shat p^2 \sin^2\theta$. The quantities
$\CF^{V_T}_{i\lambda}$ for $\lambda = L,R$ in Eq.~(\ref{eq:pippim}) are
given in terms of elements of $\Delta_0$ by
\be\label{eq:ffactors}
\CF^{V_T}_{i\lambda}(\shat) = Q_i \ts \Delta_{\gamma V_T}(\shat)
  + {2 \zeta_{i\lambda} \over{\sin 2\thw}} \Delta_{Z V_T}(\shat) \ts.
\ee
Because the $\tro$-$\tom$ mixing parameter $\epsilon_{\rho\omega}$ is
expected to be very small, the rates for $q_i \ol q_i \ra \tom \ra \pi^+_A
\pi^-_B$ are ignored here.

The cross section for $G\tpi$ production in the neutral channel is
given by
\bea\label{eq:gpineutral}
&& {d\hat\sigma(q_i \ol q_i \ra \troz,\ts \tom \ra G \tpi)
\over{d\that}} = \nn \\
&& \quad {\pi \alpha^2 \over{24 \shat}} \Biggl\{
\left(\vert\CG^{VG\tpi}_{iL}(\shat)\vert^2 +
      \vert\CG^{VG\tpi}_{iR}(\shat)\vert^2\right) \ts
      \left({\that^2 + \uhat^2 -2M^2_G M^2_{\tpi}\over{M^2_V}}\right) 
\\
&& \qquad + \left(\vert\CG^{AG\tpi}_{iL}(\shat)\vert^2 +
             \vert\CG^{AG\tpi}_{iR}(\shat)\vert^2\right) \ts
\left({\that^2 + \uhat^2 -2M^2_G M^2_{\tpi} + 4\shat
    M^2_G\over{M^2_A}}\right)\Biggr\} \ts, \nn
\eea
where, for $X = V,A$ and $\lambda = L,R$,
\be\label{eq:gfactors}
\CG^{XG\tpi} _{i\lambda} = \sum_{V_T = \troz,\tom} X_{V_T G \tpi} \CF^{V_T}
_{i\lambda} \ts.
\ee
The factor $\that^2 + \uhat^2 -2M^2_G M^2_{\tpi} = 2\shat p^2
(1+\cos^2\theta)$. The $G\tpi$ cross section in the charged channel is given
by (in the approximation of a unit CKM matrix)
\bea\label{eq:gpicharged}
& &{d\hat\sigma(u_i \ol d_i \ra \trop \ra G \tpi) \over{d\that}} =
{\pi \alpha^2 \over{48 \sin^2\thw \ts \shat}} \ts \vert\Delta_{\ts
  W\tro}(\shat)\vert^2 \\
& & \times \left\{{V^2_{\trop G\tpi}\over{M^2_V}}\biggl(\that^2 + \uhat^2
  -2M^2_G  M^2_{\tpi}\biggr) +
{A^2_{\trop G\tpi}\over{M^2_A}}\biggl(\that^2 + \uhat^2 -2M^2_G
M^2_{\tpi} + 4\shat M^2_G\biggr)\right\} \ts.\nn
\eea

The cross section for $q_i \ol q_i \ra f_j \ol f_j$ (with $m_{q_i} =
0$ and allowing $m_{f_j} \ne 0$ for $t \ol t$ production) is
\bea\label{eq:qqffrate}
{d\hat\sigma(q_i \ol q_i \ra \gamma ,\ts Z \ra \ol f_j f_j)
  \over{d\that}} &=& 
{N_f \pi \alpha^2\over{3\shat^2}} \biggl\{\left((\uhat-m_{f_j}^2)^2 +
  m_{f_j}^2\shat\right)
\ts \left(\vert\CD_{ijLL}\vert^2 + \vert\CD_{ijRR}\vert^2\right) \nn\\
&+& \left((\that-m_{f_j}^2)^2 +
  m_{f_j}^2\shat\right)\ts\left(\vert\CD_{ijLR}\vert^2 + 
  \vert\CD_{ijRL}\vert^2\right)\biggr\} \ts, 
\eea
where
\bea\label{eq:dfactors}
\CD_{ij\lambda\lambda'}(\shat) &=& Q_i Q_j \ts
\Delta_{\gamma\gamma}(\shat)  + {4\over{\sin^2 2\thw}} \ts \zeta_{i \lambda}
\ts \zeta_{\j \lambda'} \ts \Delta_{ZZ}(\shat) \\
&& + {2\over{\sin 2\thw}} \ts \biggl(\zeta_{i \lambda} Q_j
\Delta_{Z\gamma}(\shat) + Q_i \zeta_{j \lambda'} \Delta_{\gamma
  Z}(\shat)\biggr)
\ts. \nn 
\eea
Finally, the rate for the subprocess $u_i \ol d_i \ra f_j \ol f'_j$ is
\be\label{eq:udffrate}
{d\hat\sigma(u_i \ol d_i \ra W^+ \ra  f_j \ol f'_j)
  \over{d\that}} = {N_f \pi \alpha^2\over{12\sin^4\thw \ts \shat^2}} \ts
(\uhat - m^2_j)(\uhat-m^{'2}_j) \ts \vert\Delta_{WW}(\shat)\vert^2 \ts.
\ee

\section{TCSM Signal Rates at the Tevatron}

We present here a sampling of decay and production rates at the Tevatron for
$M_{\tro} = 210\,\gev$, $M_{\tom} = 200$--$220\,\gev$, $M_{\tpi} = M_{\tpipr}
= 110$ and $100\,\gev$, and $M_V = M_A = 100$--$500\,\gev$. We consider two
plausibly extreme cases for the technifermion charges, $Q_U + Q_D = 5/3$ and
$Q_U + Q_D = 0$, where $Q_D = Q_U - 1$. In the latter case $\tro$ and $\tom$
decays to $\tpi+\gamma$ are suppressed and $\tom \ra f \ol f$ decays are
forbidden altogether. In all calculations, $\Ntc = 4$ and $\sin\chi=
\sin\chipr = 1/3$. Since these calculations are at the parton level, they
should be regarded as a rough guide to what can be expected. Detailed
simulations are being carried out by Mrenna and Womersley~\cite{mwa}, who
have encoded the TCSM production and decay processes into the {\sc Pythia}
event generator~\cite{pythia} and incorporated the effects of a model
detector appropriate to Run~II conditions.

\medskip

\noindent {\underbar{{\it Case 1:} $M_{\tpi} = 110\,\gev$; $Q_U + Q_D = 5/3$}}

The total $\tro$ and $\tom$ decay rates are plotted versus $M_V$ in Fig.~1.
The dominant decay modes of $\troz$ and $\tropm$ are $W \tpi$ and $\gamma
\tpi$. The rates to these two modes are roughly equal at $M_V = M_A =
100\,\gev$, but the $\gamma \tpi$ rate falls off as $M_V^{-2}$. The total
widths are about $1\,\gev$ with a partial width to all fermion pairs, $f \ol
f$, of about $30\,\mev$. At $M_V = 100\,\gev$, the width of a 200 (220)~GeV
$\tom$ to $\gamma \tpi$ is 300 (560)~MeV and its ($M_V$-independent) width to
$f \ol f$ is $45\,\mev$. The rapid fall of $\Gamma(\tom)$ with $M_V$ is
apparent. At $M_V = 300\,\gev$, the $\tom$'s branching fraction to $f \ol f$
is already 55\%.~\footnote{These decay rates are calculated from the formulas
of Section~2. They ignore the effects of mixing, which are not entirely
negligible for $\troz$ and $\tom$. Nevertheless, they give a fair estimate of
the relative contributions of the resonances to individual final state
production rates.}

In Fig.~2 we show the total $\gamma \tpi$ production rate ($\gamma \tpiz$,
$\gamma \tpipr$, and $\gamma \tpipm$) as a function of $M_V$ for various
$M_{\tom}$. Again, the rapid fall with increasing $M_V$ is apparent, with the
cross sections dropping from $5\,\pb$ to $1\,\pb$. The dependence on the
input $\tro$--$\tom$ mass difference is not significant over the range we
considered. Due to the additional $\gamma\tpi$ channels, this rate at $M_V =
100\,\gev$ is twice what we found in Ref.~\cite{elw} where we considered only
$\tom \ra \gamma \tpiz$. Note that our calculations are done in lowest order
QCD with EHLQ Set~1 parton distribution functions~\cite{ehlq}. For these
Drell-Yan processes, next-order corrections to the cross sections and the
distribution functions would increase the rates by about 50\% ($K \simeq
1.5$). Thus, assuming that $\tpiz$ and $\tpipr$ decay mainly to $b \ol b$ and
$\tpip$ to $c \ol b$, we expect that Run~II searches for $\gamma$ plus two
jets with a single $b$-tag can cover the range $|Q_U + Q_D| \simle 1$ up to
$M_{V_T} \simeq 350\,\gev$. It is also important to look for the $\tpipr$ in
its two-gluon decay mode. It is an open question whether this could be seen
above the $\gamma$ plus two untagged jets background for, say, $B(\tpiz \ra
gg) = 0.5$.

Figure~3 shows the $W\tpi$ and $Z\tpi$ production rates.~\footnote{The $W_L
W_L$ and $W_L Z_L$ cross sections, suppressed by $\sin^4\chi$, are less than
$0.5\,\pb$, not large enough to see above the standard model backgrounds.}
The $W\tpi$ cross sections add up to 4--5~pb without the $K$-factor, for all
the inputs of this case. This is about the same found in Ref.~\cite{elw} even
though there has been a doubling of the $\gamma\tpi$ rate for $M_V =
100\,\gev$. The reason for this is the new contribution from the transversely
polarized $W_T \tpi$ mode; see Eq.(\ref{eq:longdecay}) and Table~1. We expect
that, so long as $M_{\tro} \simge M_W + M_{\tpi}$, the process $\tro \ra
W\tpi$ could be observed up to $M_{\tro} \simeq 400\,\gev$ in Run~II. Unless
there is substantial $\tpiz$--$\tpipr$ mixing, very little of the $W\tpi$
rate involves the isosinglet $\tpipr$. To test for this mixing, one can look
for $\tpipr \ra gg$ in association with a $W$. Such a signal should be
discernible above background if the cross section is a few
picobarns~\cite{mwb}.

The $Z \tpi$ rate is only about $0.9\,\pb$ for $M_{\tro} = 210\,\gev$, about
50\% less than we found in the simple model employed in Ref.~\cite{elw}. If
the $\tro$ and $\tpi$ are discovered in any of their larger-rate channels, it
would be interesting to confirm them here. At this cross section, it may just
be possible to detect $\ell^+ \ell^- jj$ with a $b$-tag in $2\,\ifb$ of
data. Another interesting and challenging signature is $\etmiss$ plus two
jets with a $b$-tag arising from $Z\tpi \ra \nu \ol \nu b j$.

Finally, we also show in Fig.~3 the total $\tpi\tpi$ cross section for
$M_{\tro} = 210\,\gev$ and $M_{\tpi} = 110\,\gev$. This continuum production
rate is only $0.12\,\pb$. Even with very efficient $b$-identification, it
seems unlikely that it will be possible to detect technipions in this mode
above the four-jet background.

Technivector decays to lepton pairs may be an accessible signature at the
Tevatron. Figures~4 and~5 show the mass distribution, $d\sigma(p \ol p \ra
e^+ e^-)/d\rshat$, for the extreme cases $M_V = 100$ and $500\,\gev$. The
input $\tro$--$\tom$ mass splittings in each figure are zero and $\pm
10\,\gev$. From this, one can judge the effect of mixing. For all $M_V$, most
of the signal comes from the $\tom$ because it is proportional to $(Q_U +
Q_D)^2 = 25/9$ and its branching ratio to $e^+ e^-$ is several times larger
than the $\troz$'s. The signal-plus-background rates for $M_V = 100\,\gev$,
integrated over the entire resonance region from $195$ to $225\,\gev$, are
0.19, 0.17, and $0.15\,\pb$ for $M_{\tom} = 200$, 210, and $220\,\gev$, while
the standard-model background is $0.13\,\pb$. For $M_V = 500\,\gev$, the
branching ratio of $\tom$ to $e^+e^-$ increases by a factor of~7 and the
total $e^+e^-$ rate doubles to 0.38, 0.30, and $0.31\,\pb$. No smearing due
to detector resolution was included here. The separated resonances are just
at or below the detectors' dielectron mass resolutions. It will be
interesting to see what these mass distributions look like when the effects
of a real detector are included~\cite{mn}.

There is no observable $\tropm$ enhancement in the $\ell^\pm \nu_\ell$ cross
section. This is clear from the (theoretical) invariant mass distributions of
Fig.~6. The signal rate is small because $B(\tropm \ra \ell^\pm\nu_\ell)$ is.
This is true for all the input parameters we considered.

\medskip

\noindent {\underbar{{\it Case 2:} $M_{\tpi} = 110\,\gev$; $Q_U + Q_D = 0$}}

The sharp decrease in the $\tom \ra \gamma\tpi$ and $e^+e^-$ rates when $Q_U
+ Q_D = 0$ is apparent in Figs.~7--9. Because most of the $\gamma\tpi$ cross
section in case~1 come from $\tom$ production, it has in this case fallen by
a factor of 20--50, depending on $M_V$. The $e^+e^-$ signal rate is tiny
because it all comes from $\troz$.  Finally, because $\tom$ mixing with
$\gamma$ and $Z$ vanishes when $Q_U + Q_D = 0$ (see Eq.~(\ref{eq:fgv})), so
does $\tro$--$\tom$ mixing, and all the production rates are independent of
$M_{\tom}$. The $W\tpi$ cross section is still large, about $4\,\pb$, and
represents the best way to discover $\tro$ and $\tpi$ in this extreme
case. The $\tpi\tpi$ rate is still about $0.12\,\pb$.

\medskip

\noindent {\underbar{{\it Case 3:} $M_{\tpi} = 100\,\gev$; $Q_U + Q_D = 5/3$}}

Now, the $\tro$ is just above threshold to decay into a pair of
technipions. The $\tro$ widths have increased to 2--$3\,\gev$; see
Fig.~10. This has caused a 25\% decrease in the $\gamma\tpi$ rates compared
to case~1 (Fig.~11), but this signal is still a relatively easy one in Run~II
up to $M_{\tom} \simeq 350\,\gev$. Figure~12 shows the $W\tpi$, $Z\tpi$ and
$\tpi\tpi$ cross sections versus $M_V$. The $W\tpi$ rate is $3\,\pb$, still
large enough to detect, and $\sigma(Z\tpi) =0.5$--$1.0\,\pb$, as before. We
still expect that $\tro \ra W\tpi$ could be detected in Run~II up to
$M_{\tro} \simeq 400\,\gev$ so long as $M_{\tro} \simge M_W + M_{\tpi}$. The
$\tpi\tpi$ rate has grown a factor of 20--30 to 2.5--4~pb because it is
unsuppressed by powers of $\sin\chi$. Roughly half this is $\tpip\tpim \ra c
\ol b b \ol c$ and half is $\tpipm \tpiz \ra c \ol b b \ol b$. It should be
possible to see such signals at rates this large in Run~II. The ultimate mass
reaches for $\tro \ra \tpi\tpi$ in Run~II and in the proposed
$20$--$30\,\ifb$ Run~III remain to be determined by detailed
simulations. Finally, as we see in Figs.~13 and~14, the $e^+e^-$ rates again
are due mainly to $\tom$ production and little affected by the lowered $\tpi$
mass. Integrated over the resonance region, they are very similar to those
found in case~1: signal-plus-background rates of 0.18, 0.16, and $0.14\,\pb$
over a background of $0.13\,\pb$ for $M_V = 100\,\gev$; they are 0.36, 0.24,
and $0.29\,\pb$ for $M_V = 500\,\gev$.

\section{Concluding Remarks}

The straw-man model studied in this paper assumes a relatively uncluttered,
minimal spectrum for low-scale technicolor. We believe that the parameters
chosen for study are sufficiently generic to warrant our expectation that, up
to $M_{\tro} \simeq 400\,\gev$, such a spectrum can be ruled out---or
established---in Run~II at the Tevatron. A richer and more complicated
spectrum, due to several low-lying technifermion doublets might be more
representative of low-scale technicolor and might be more (or less) difficult
to pin down experimentally. One generalization of the TCSM would include a
minimal set of $SU(3)$-triplet technifermion doublets. We plan to carry it
out in the near future. Together with the color-singlet states discussed
here, that would make for a very rich experimental program in technicolor,
even into the proposed Tevatron Run~III.

\section{Acknowledgements}

I am grateful for inspiration, advice and encouragement to R.~S.~Chivukula,
E.~Eichten, U.~Heintz, S.~Mrenna, M.~Narain, S.~Parke, J.~Womersley, and
other members of the ``Strong Dynamics for Run~~II Workshop'' at Fermilab. I
thank the Aspen Center for Physics and Fermilab for their hospitality during
various stages of this work. This research was supported in part by the
Department of Energy under Grant~No.~DE--FG02--91ER40676.

\vfil\eject

\begin{figure}[tb]
\vbox to 8cm{
\vfill
\includegraphics{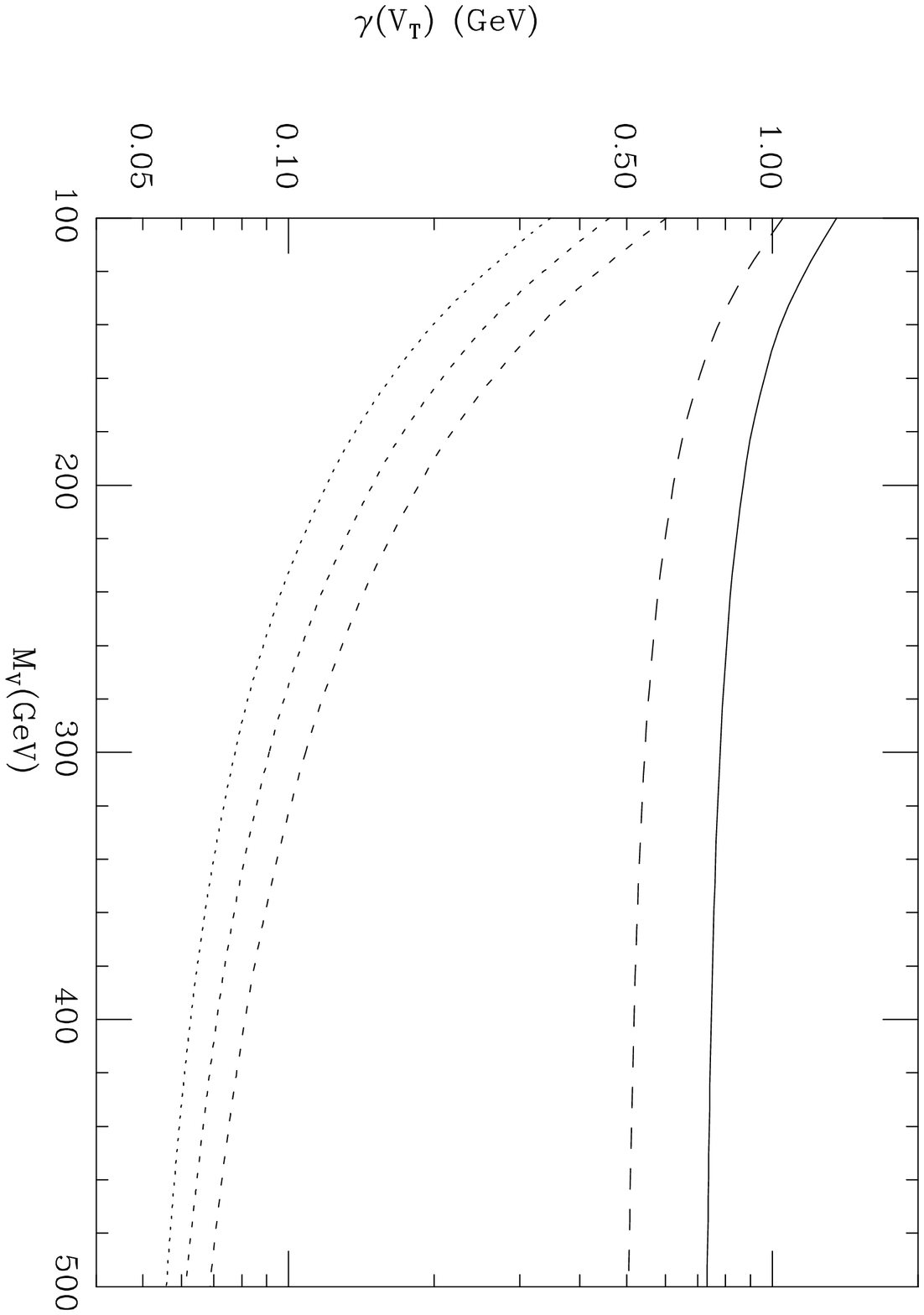}
\vfill
}
\caption{Technivector meson decay rates versus $M_V = M_A$ for $\troz$ (solid
  curve) and $\tropm$ (long-dashed) with $M_{\tro} = 210\,\gev$, and $\tom$
  with $M_{\tom} = 200$ (lower dotted), 210 (lower short-dashed), and
  $220\,\gev$ (lower medium-dashed); $Q_U + Q_D = 5/3$ and $M_{\tpi} =
  110\,\gev$.
\label{fig:a}}
\end{figure}

\vfil\eject

\begin{figure}[tb]
\vbox to 8cm{
\vfill
\includegraphics{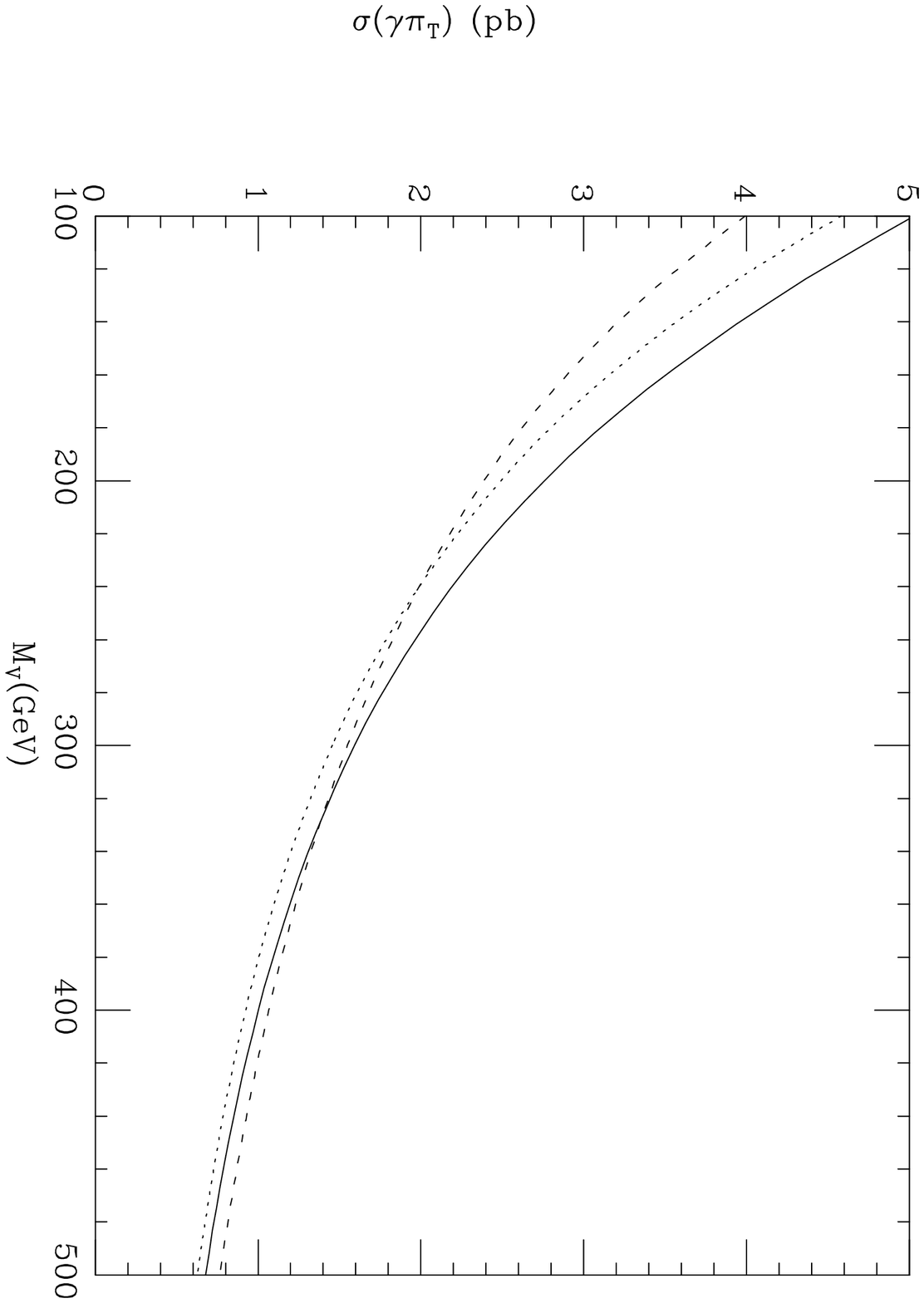}
\vfill
}
\caption{Production rates in $p \ol p$ collisions at $\ecm = 2\,\tev$
for the sum of $\tom$, $\troz$, $\tropm \ra \gamma
  \tpi$ versus $M_V$, for $M_{\tro} = 210\,\gev$ and $M_{\tom} = 200$ (dotted
  curve), 210 (solid), and $220\,\gev$ (short-dashed); $Q_U + Q_D = 5/3$, and
  $M_{\tpi} = 110\,\gev$.
\label{fig:b}}
\end{figure}

\vfil\eject

\begin{figure}[tb]
\vbox to 8cm{
\vfill
\includegraphics{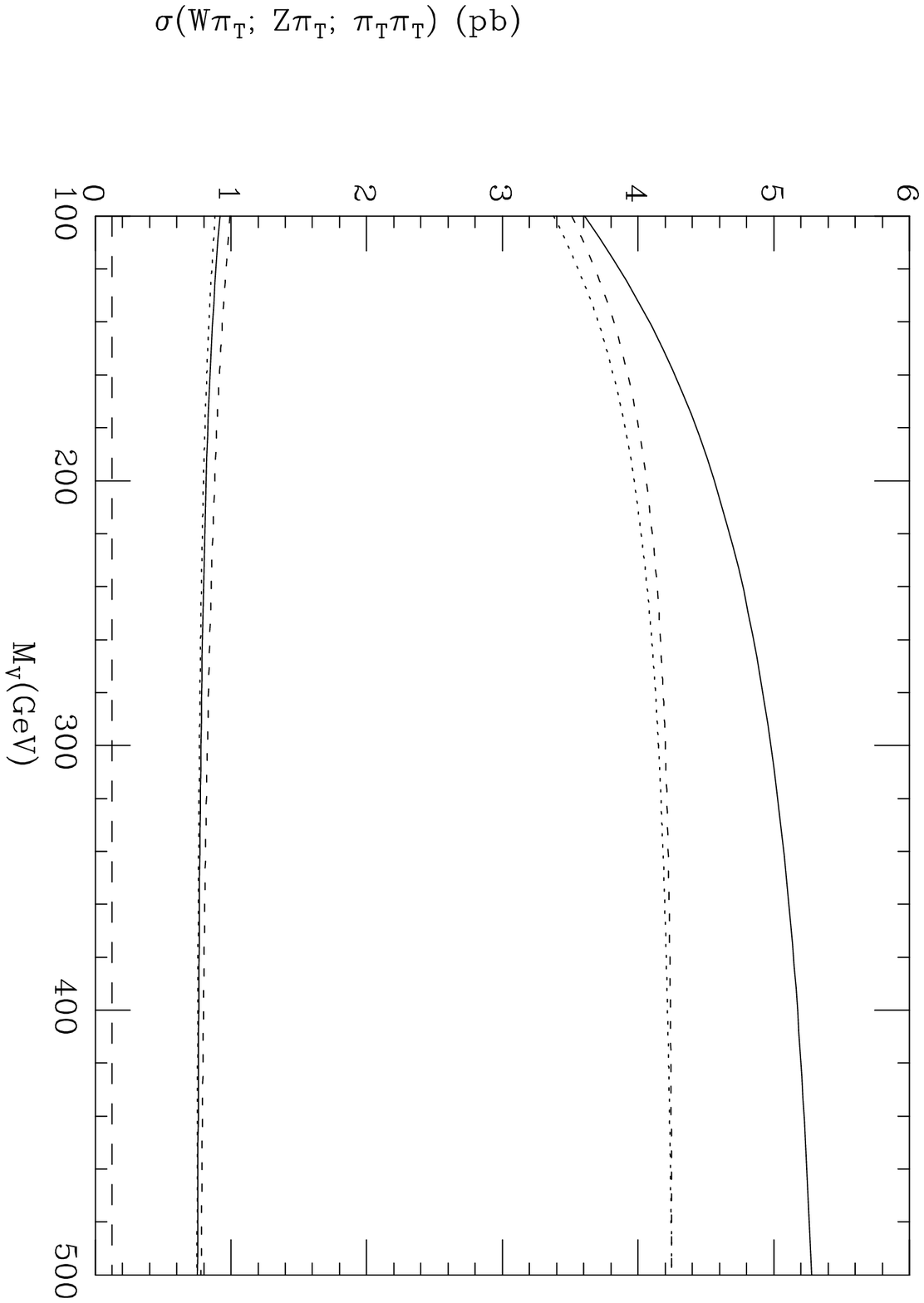}
\vfill
}
\caption{Production rates for $\tom$, $\troz$, $\tropm \ra W \tpi$ (upper
  curves) and $Z\tpi$ (lower curves) versus $M_V$, for $M_{\tro} = 210\,\gev$
  and $M_{\tom} = 200$ (dotted curve), 210 (solid), and $220\,\gev$
  (short-dashed); $Q_U + Q_D = 5/3$ and $M_{\tpi} = 110\,\gev$. Also shown
  is $\sigma(\tro \ra \tpi\tpi)$ (lowest dashed curve).
\label{fig:c}}
\end{figure}

\vfil\eject

\begin{figure}[tb]
\vbox to 8cm{
\vfill
\includegraphics{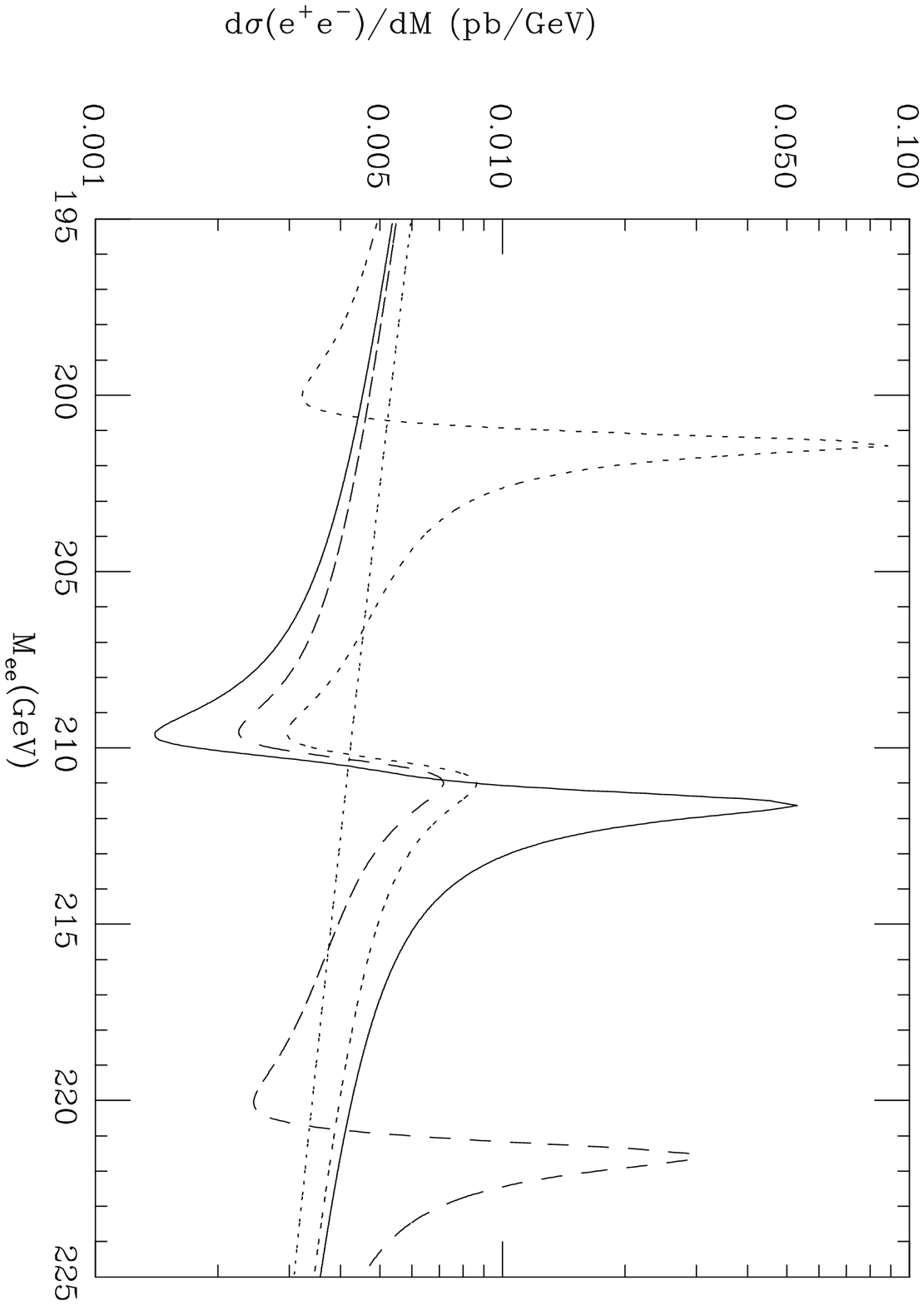}
\vfill
}
\caption{Invariant mass distributions for $\tom$, $\troz \ra e^+e^-$
  for $M_{\tro} = 210\,\gev$ and $M_{\tom} = 200$ (short-dashed curve), 210
  (solid), and $220\,\gev$ (long-dashed); $M_V = 100\,\gev$. The standard
  model background is the sloping dotted line. $Q_U + Q_D = 5/3$ and
  $M_{\tpi} = 110\,\gev$.
\label{fig:d}}
\end{figure}

\vfil\eject

\begin{figure}[tb]
\vbox to 8cm{
\vfill
\includegraphics{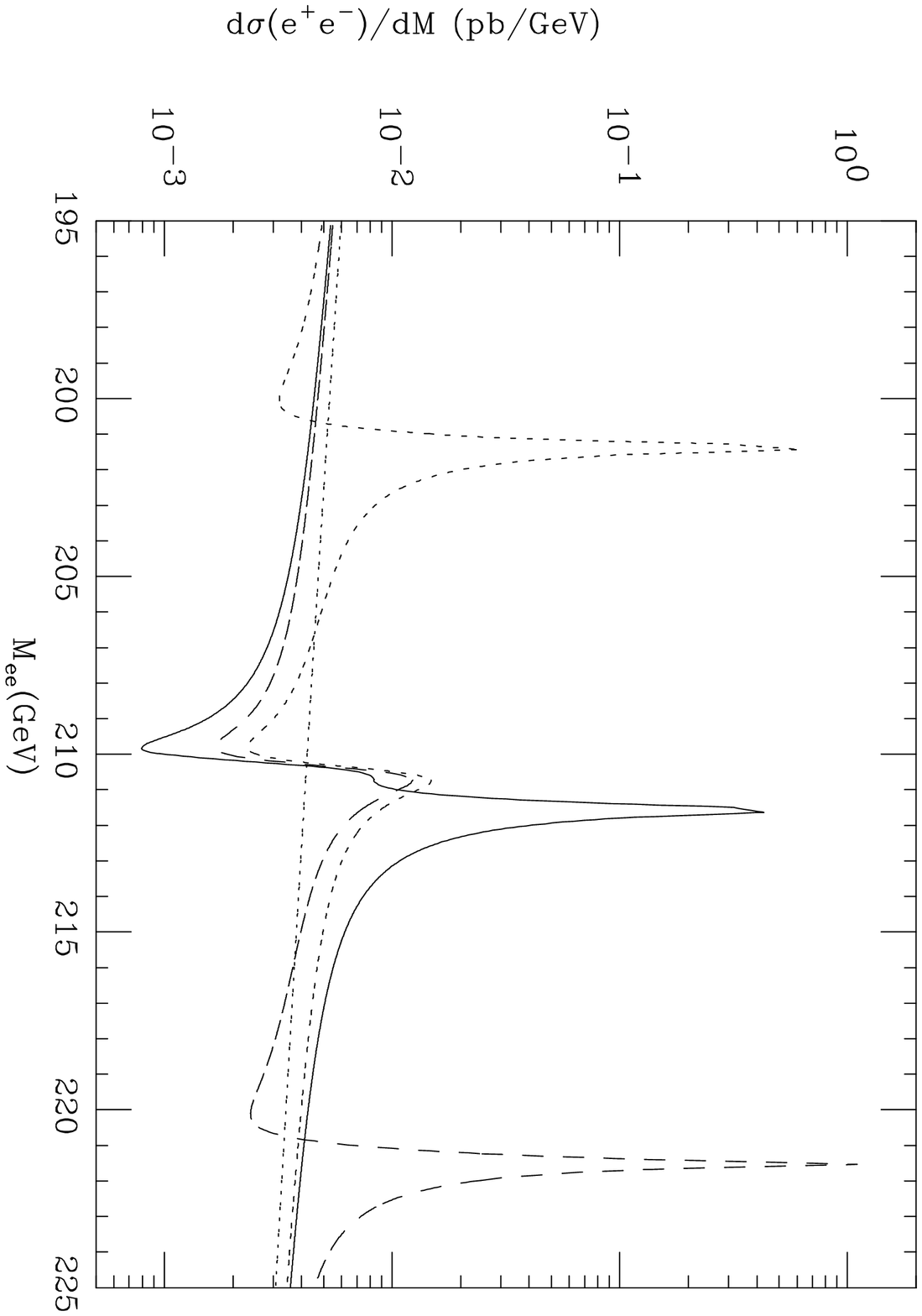}
\vfill
}
\caption{Invariant mass distributions for $\tom$, $\troz \ra e^+e^-$
  for $M_{\tro} = 210\,\gev$ and $M_{\tom} = 200$ (short-dashed curve), 210
  (solid), and $220\,\gev$ (long-dashed); $M_V = 500\,\gev$. The standard
  model background is the sloping dotted line. $Q_U + Q_D = 5/3$ and
  $M_{\tpi} = 110\,\gev$.
\label{fig:e}}
\end{figure}

\vfil\eject

\begin{figure}[tb]
\vbox to 8cm{
\vfill
\includegraphics{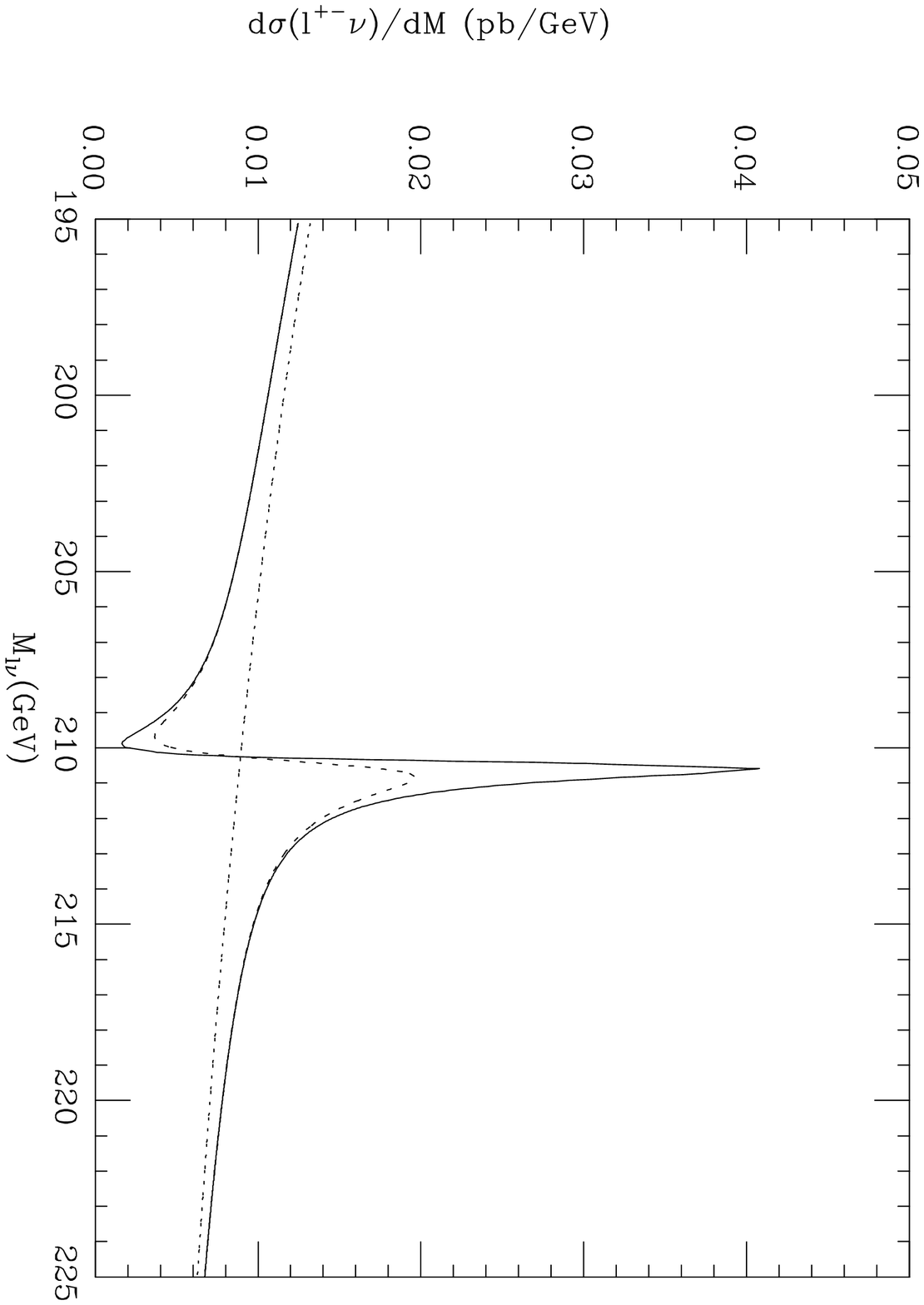}
\vfill
}
\caption{Invariant mass distributions for $\tropm \ra \ell^\pm \nu$
  for $M_{\tro} = 210\,\gev$ and $M_V= 100\,\gev$ (dashed curve) and
  $500\,\gev$ (solid); The standard model background is the sloping dotted
  line. $Q_U + Q_D = 5/3$ and $M_{\tpi} = 110\,\gev$.
\label{fig:f}}
\end{figure}

\vfil\eject

\begin{figure}[tb]
\vbox to 8cm{
\vfill
\includegraphics{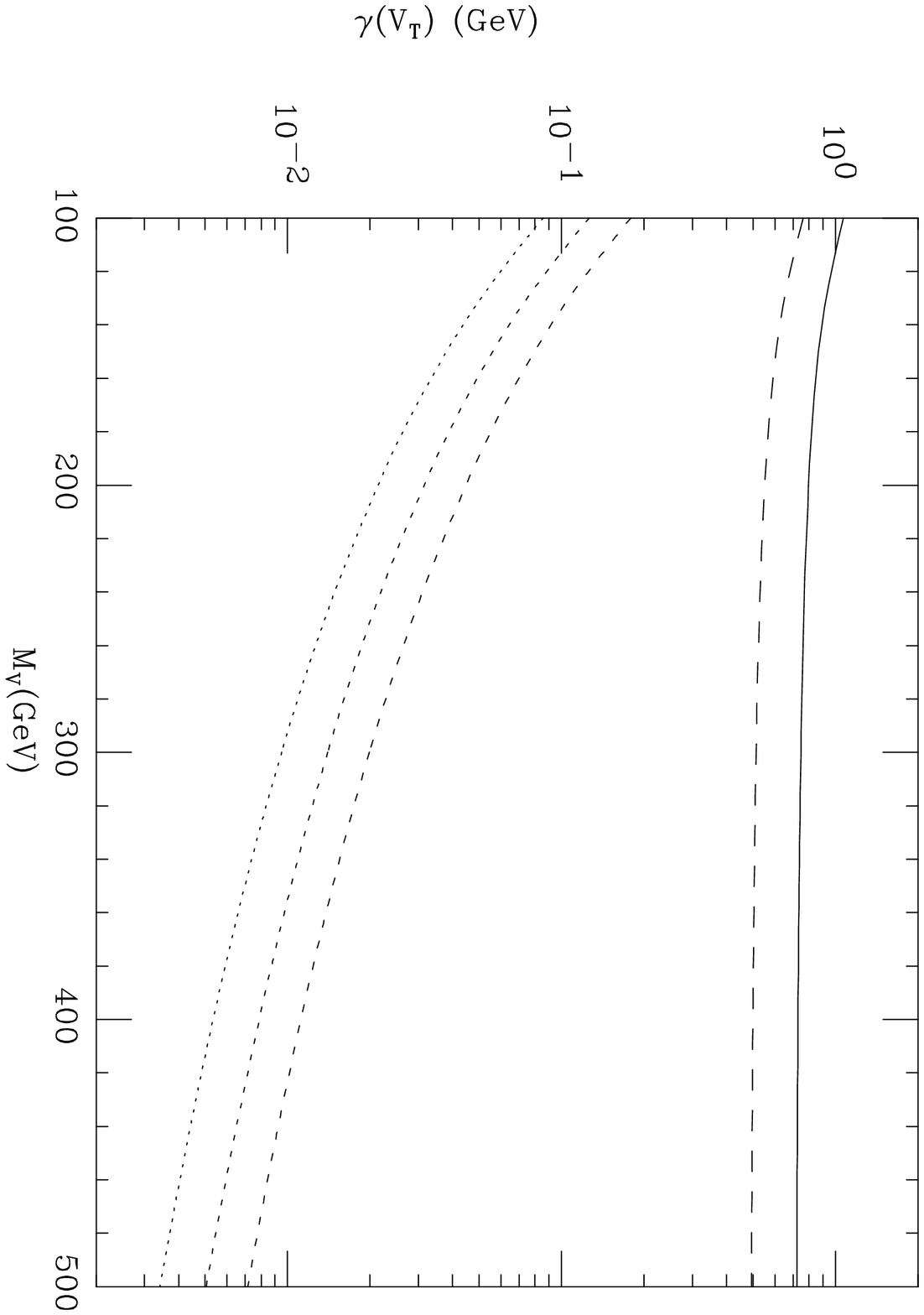}
\vfill
}
\caption{Technivector meson decay rates versus $M_V = M_A$ for $\troz$ (solid
  curve) and $\tropm$ (long-dashed) with $M_{\tro} = 210\,\gev$, and $\tom$
  with $M_{\tom} = 200$ (lower dotted), 210 (lower short-dashed), and
  $220\,\gev$ (lower medium-dashed); $Q_U + Q_D = 0$ and $M_{\tpi} =
  110\,\gev$.
\label{fig:g}}
\end{figure}

\vfil\eject

\begin{figure}[tb]
\vbox to 8cm{
\vfill
\includegraphics{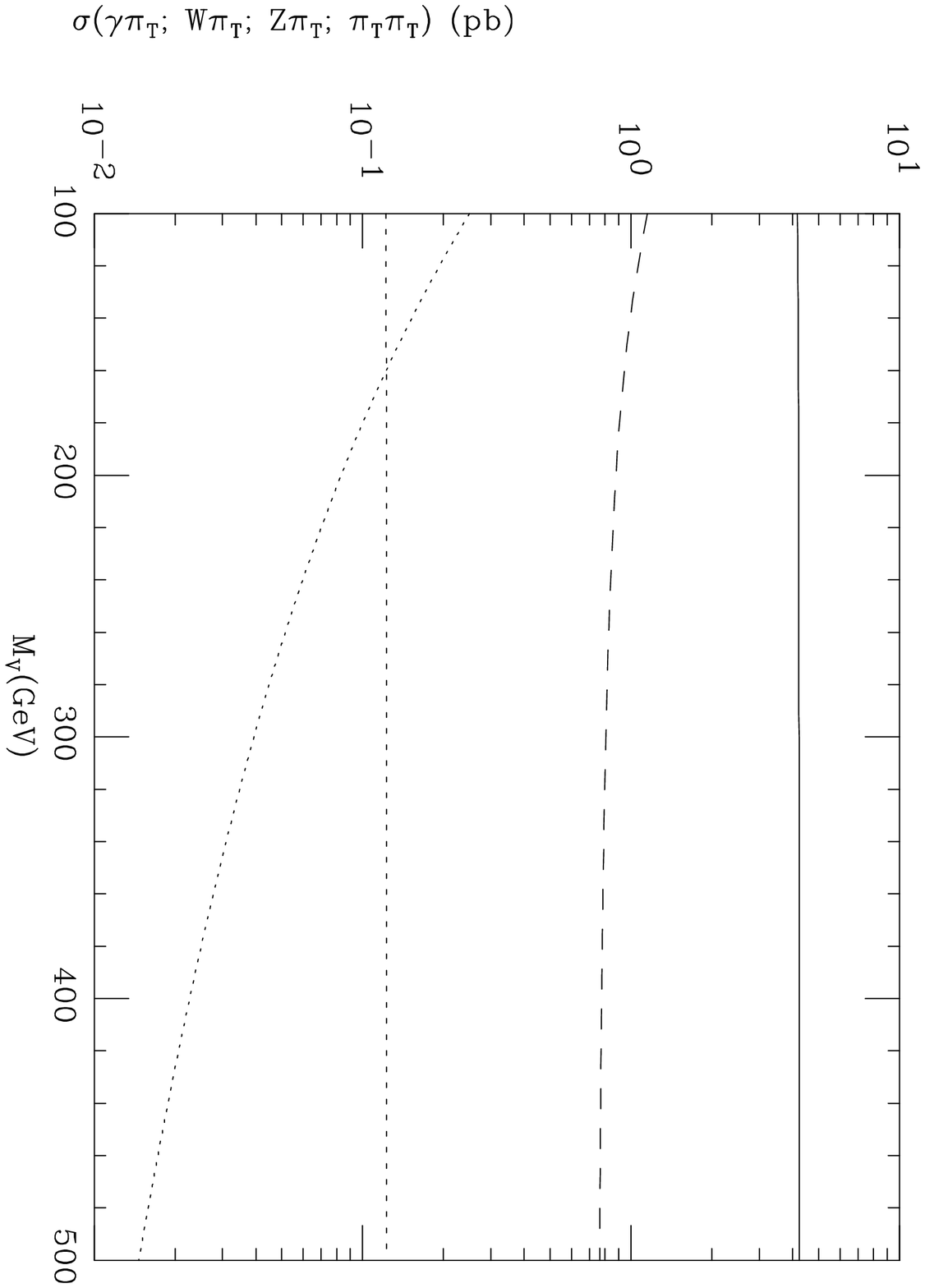}
\vfill
}
\caption{Production rates for $\tom$, $\troz$, $\tropm \ra W \tpi$ (solid
  curve), $Z\tpi$ (long-dashed), $\tpi\tpi$ (short-dashed) and $\gamma\tpi$
  (dotted) versus $M_V$, for $M_{\tro} = 210\,\gev$ and $M_{\tom} =
  200$--$220\,\gev$; $Q_U + Q_D = 0$ and $M_{\tpi} = 110\,\gev$.
\label{fig:h}}
\end{figure}

\vfil\eject

\begin{figure}[tb]
\vbox to 8cm{
\vfill
\includegraphics{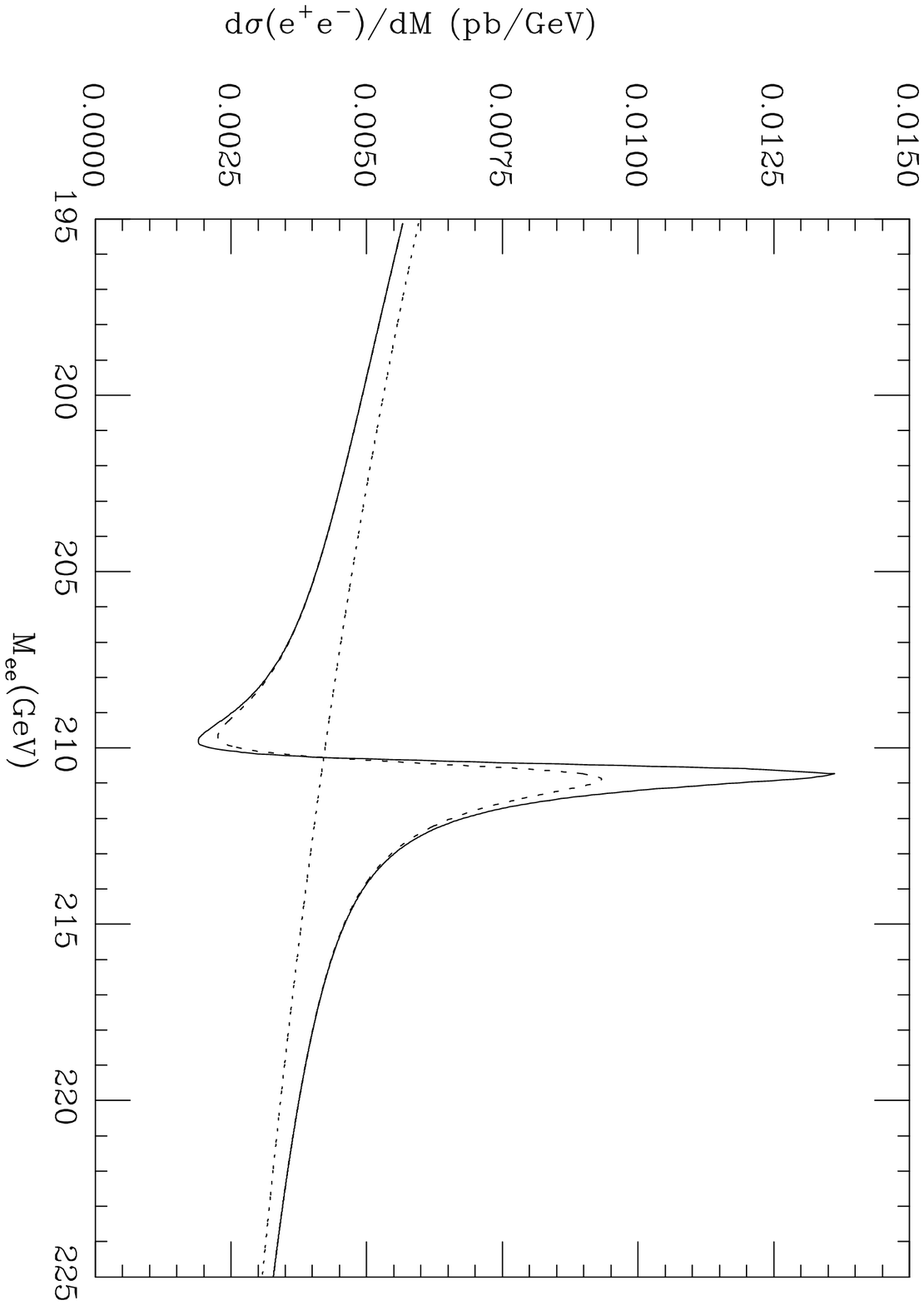}
\vfill
}
\caption{Invariant mass distributions for $\troz \ra e^+e^-$
  for $M_{\tro} = 210\,\gev$; $M_V = 100\,\gev$ (dashed curve) and
  $500\,\gev$ (solid). The standard model background is the sloping dotted
  line. $Q_U + Q_D = 0$ and $M_{\tpi} = 110\,\gev$.
\label{fig:i}}
\end{figure}

\vfil\eject

\begin{figure}[tb]
\vbox to 8cm{
\vfill
\includegraphics{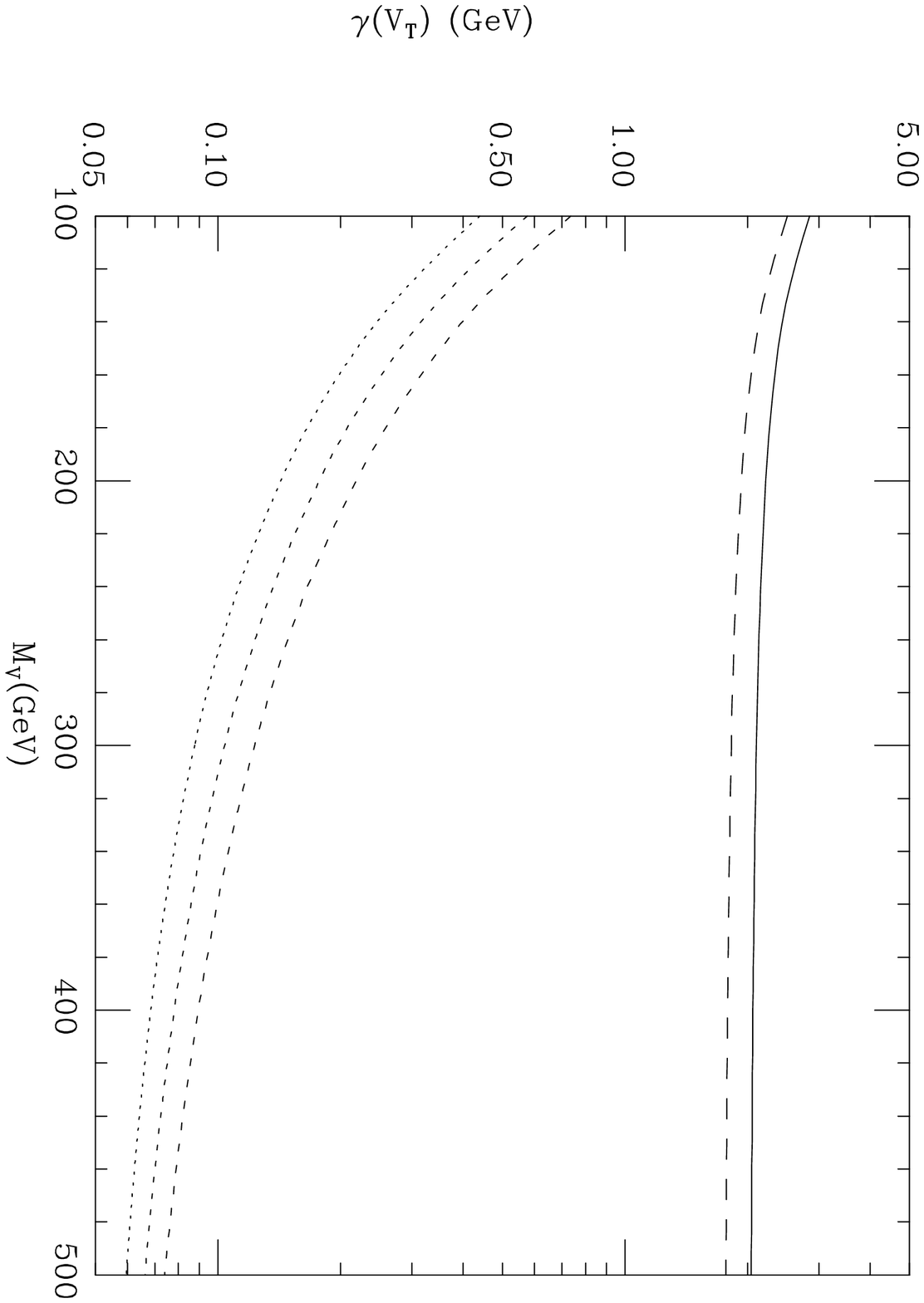}
\vfill
}
\caption{Technivector meson decay rates versus $M_V = M_A$ for $\troz$ (solid
  curve) and $\tropm$ (long-dashed) with $M_{\tro} = 210\,\gev$, and $\tom$
  with $M_{\tom} = 200$ (lower dotted), 210 (lower short-dashed), and
  $220\,\gev$ (lower medium-dashed); $Q_U + Q_D = 5/3$ and $M_{\tpi} =
  100\,\gev$.
\label{fig:j}}
\end{figure}

\vfil\eject

\begin{figure}[tb]
\vbox to 8cm{
\vfill
\includegraphics{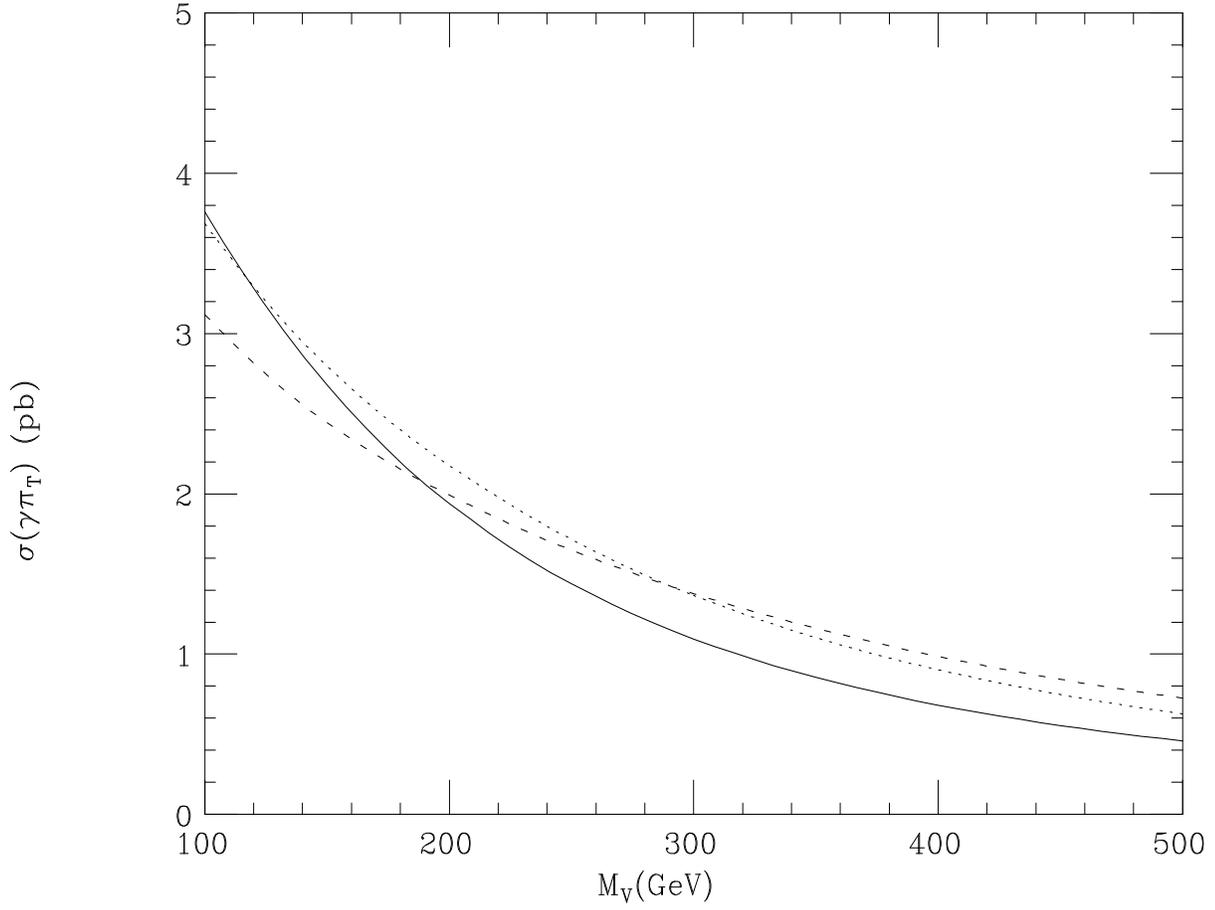}
\vfill
}
\caption{Production rates for the sum of $\tom$, $\troz$, $\tropm \ra \gamma
  \tpi$ versus $M_V$, for $M_{\tro} = 210\,\gev$ and $M_{\tom} = 200$ (dotted
  curve), 210 (solid), and $220\,\gev$ (short-dashed); $Q_U + Q_D = 5/3$, and
  $M_{\tpi} = 100\,\gev$.
\label{fig:k}}
\end{figure}

\vfil\eject

\begin{figure}[tb]
\vbox to 8cm{
\vfill
\includegraphics{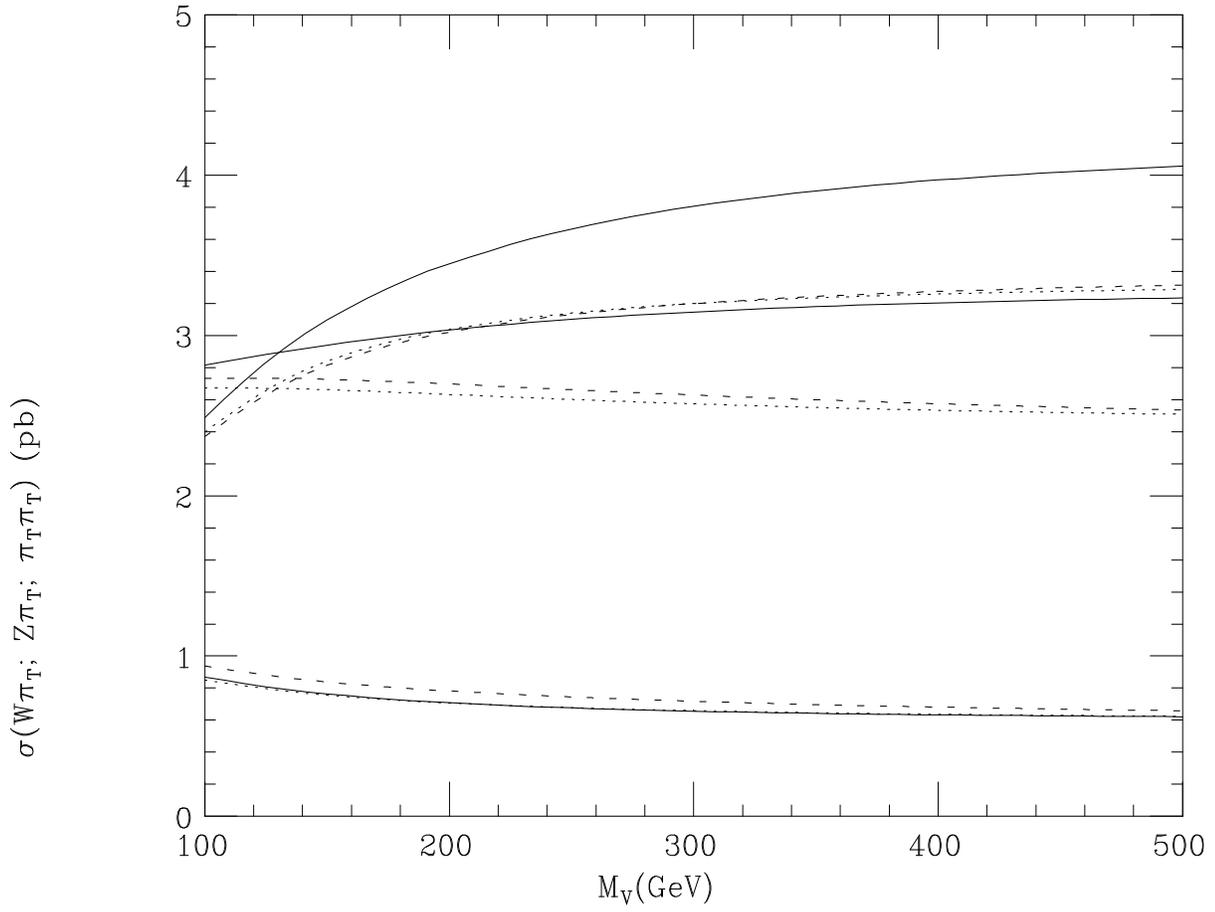}
\vfill
}
\caption{Production rates for $\troz$, $\tropm$, $\tom \ra \tpi\tpi$ (upper
  three curves), $W\tpi$ (middle three curves), and $Z\tpi$ (lower curves)
  versus $M_V$, for $M_{\tro} = 210\,\gev$ and $M_{\tom} = 200$ (dotted), 210
  (dashed), and $220\,\gev$ (solid); $Q_U + Q_D = 5/3$ and $M_{\tpi} =
  100\,\gev$.
\label{fig:l}}
\end{figure}

\vfil\eject

\begin{figure}[tb]
\vbox to 8cm{
\vfill
\includegraphics{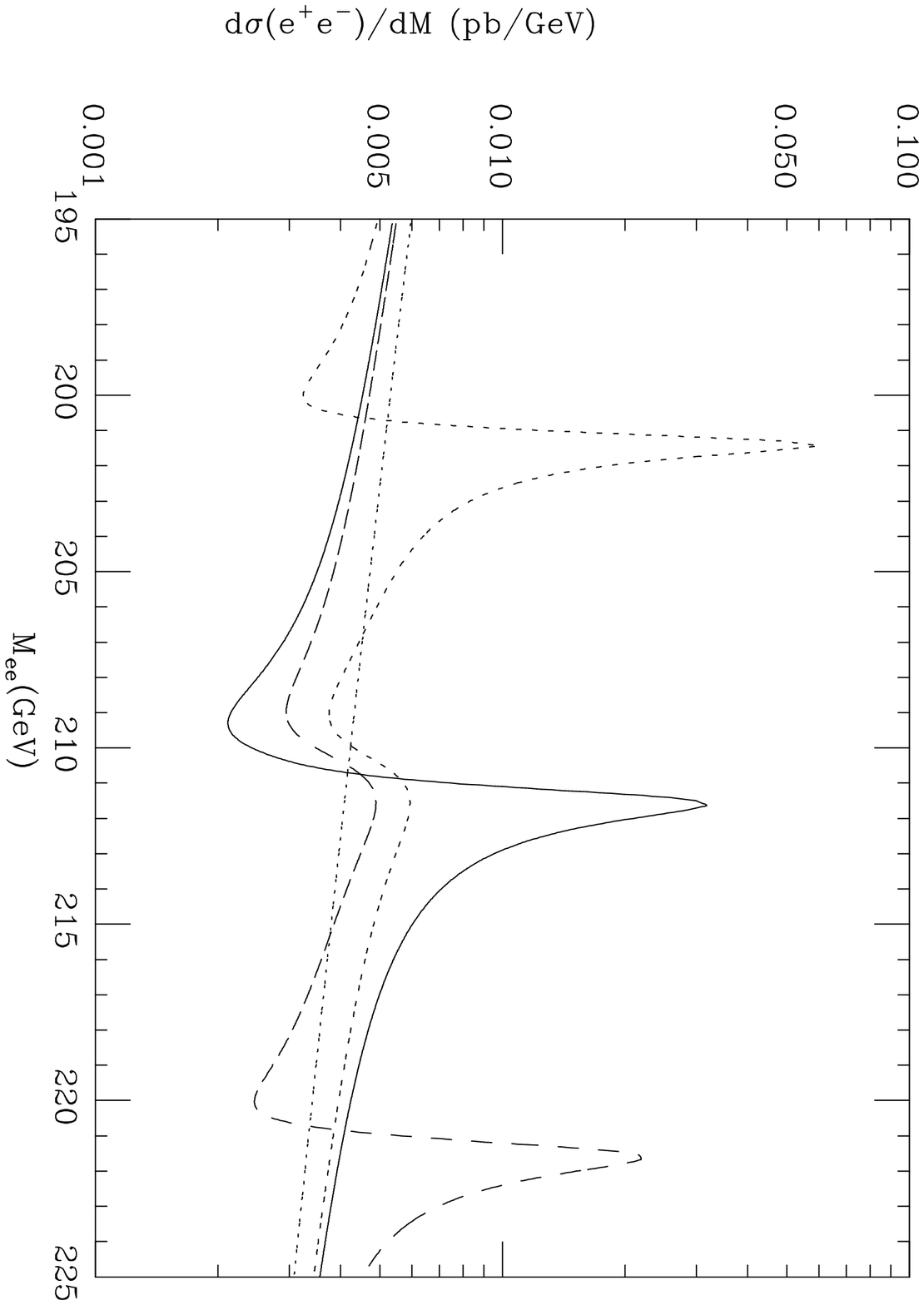}
\vfill
}
\caption{Invariant mass distributions for $\tom$, $\troz \ra e^+e^-$
  for $M_{\tro} = 210\,\gev$ and $M_{\tom} = 200$ (short-dashed curve), 210
  (solid), and $220\,\gev$ (long-dashed); $M_V = 100\,\gev$. The standard
  model background is the sloping dotted line. $Q_U + Q_D = 5/3$ and
  $M_{\tpi} = 100\,\gev$.
\label{fig:m}}
\end{figure}

\vfil\eject

\begin{figure}[tb]
\vbox to 8cm{
\vfill
\includegraphics{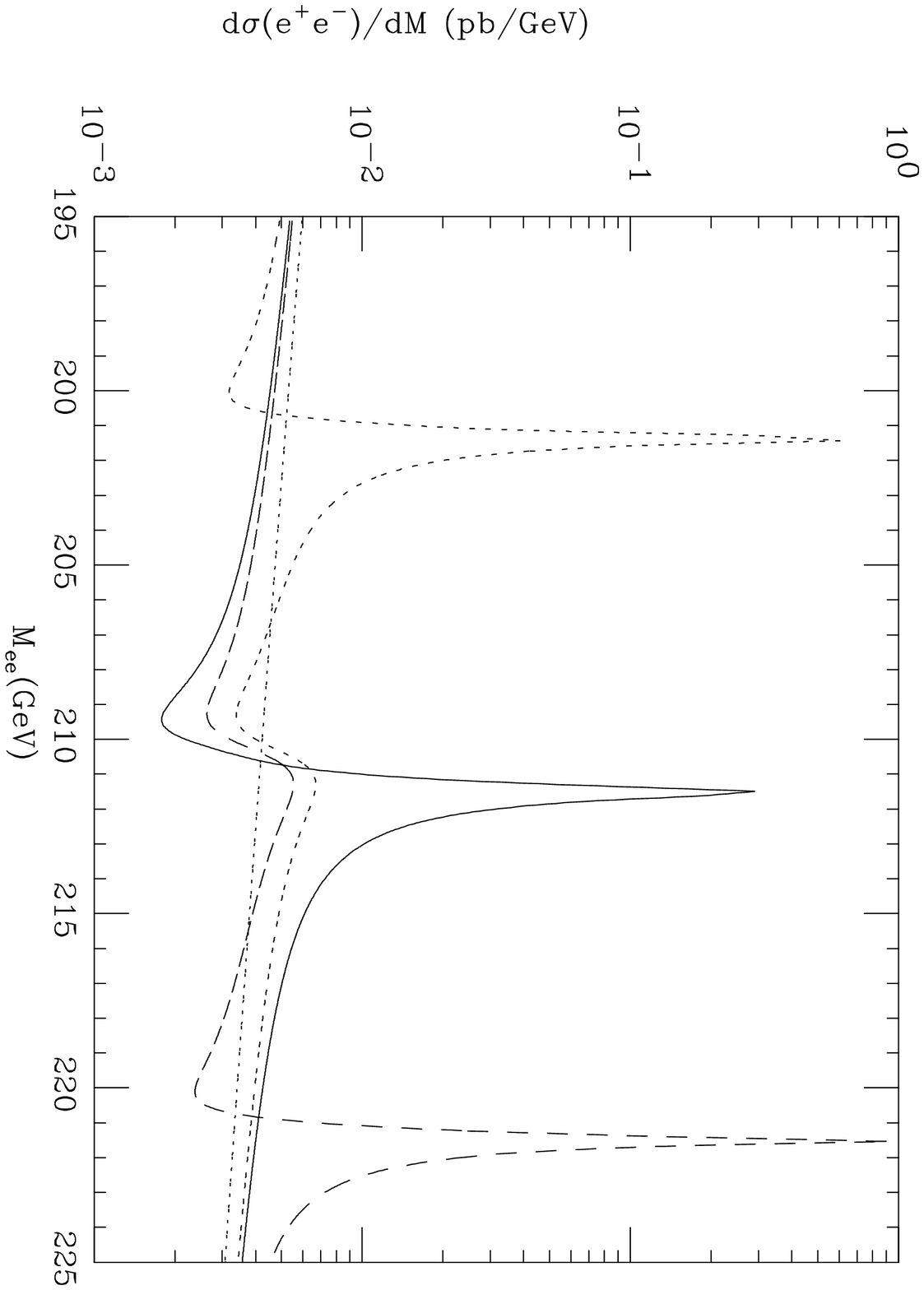}
\vfill
}
\caption{Invariant mass distributions for $\tom$, $\troz \ra e^+e^-$
  for $M_{\tro} = 210\,\gev$ and $M_{\tom} = 200$ (short-dashed curve), 210
  (solid), and $220\,\gev$ (long-dashed); $M_V = 500\,\gev$. The standard
  model background is the sloping dotted line. $Q_U + Q_D = 5/3$ and
  $M_{\tpi} = 100\,\gev$.
\label{fig:n}}
\end{figure}


\begin{thebibliography}{99}
%
\bibitem{wtc}B.~Holdom, Phys.~Rev.~{\bf D24}, 1441 (1981);
Phys.~Lett.~{\bf 150B}, 301 (1985);
T.~Appelquist, D.~Karabali and L.~C.~R. Wijewardhana,
Phys.~Rev.~Lett.~{\bf 57}, 957 (1986);
T.~Appelquist and L.~C.~R.~Wijewardhana, Phys.~Rev.~{\bf D36}, 568
(1987);
K.~Yamawaki, M.~Bando and K.~Matumoto, Phys.~Rev.~Lett.~{\bf 56}, 1335
(1986);
T.~Akiba and T.~Yanagida, Phys.~Lett.~{\bf 169B}, 432 (1986).
%
%
\bibitem{topcondref}Y.~Nambu, in {\it New Theories in Physics}, Proceedings of
the XI International Symposium on Elementary Particle Physics, Kazimierz,
Poland, 1988, edited by Z.~Adjuk, S.~Pokorski and A.~Trautmann (World
Scientific, Singapore, 1989); Enrico Fermi Institute Report EFI~89-08
(unpublished);
V.~A.~Miransky, M.~Tanabashi and K.~Yamawaki, Phys.~Lett.~{\bf
221B}, 177 (1989); Mod.~Phys.~Lett.~{\bf A4}, 1043 (1989);
W.~A.~Bardeen, C.~T.~Hill and M.~Lindner, Phys.~Rev.~{\bf D41},
1647 (1990).
C.~T. Hill, Phys.~Lett.~{\bf 266B}, 419 (1991) ;
S.~P.~Martin, Phys.~Rev.~{\bf D45}, 4283 (1992);
{\it ibid}~{\bf D46}, 2197 (1992); Nucl.~Phys.~{\bf B398}, 359 (1993);
M.~Lindner and D.~Ross, Nucl.~Phys.~{\bf  B370}, 30 (1992);
R.~B\"{o}nisch, Phys.~Lett.~{\bf 268B}, 394 (1991);
C.~T.~Hill, D.~Kennedy, T.~Onogi, H.~L.~Yu, Phys.~Rev.~{\bf D47}, 2940 
(1993).
%
%
\bibitem{tctwohill}C.~T.~Hill, Phys.~Lett.~{\bf 345B}, 483 (1995).
%
%
\bibitem{tctwoklee}K.~Lane and E.~Eichten, Phys.~Lett.~{\bf B352}, 382
(1995) ;
K.~Lane, Phys.~Rev.~{\bf D54}, 2204 (1996);
K.~Lane, Phys.~Lett.~{\bf B433}, 96 (1998).
%
%
\bibitem{tc} S.~Weinberg, Phys.~Rev.~{\bf D19}, 1277 (1979);
L.~Susskind, Phys.~Rev.~{\bf D20}, 2619 (1979).
%
%
\bibitem{etc} E.~Eichten and K.~Lane, Phys.~Lett.~{\bf B90}, 125 (1980).
%
%
\bibitem{multi}K.~Lane and E.~Eichten, Phys. Lett. {\bf B222}, 274 (1989);
K.~Lane and M.~V.~Ramana, Phys.~Rev.~{\bf D44}, 2678 (1991).
%
%
\bibitem{elw} E.~Eichten and K.~Lane, Phys.~Lett.~{\bf B388}, 803 (1996);
E.~Eichten, K.~Lane and J.~Womersley, Phys.~Lett.~{\bf B405}, 305 (1997);
E.~Eichten, K.~Lane and J.~Womersley, Phys.~Rev.~Lett.~{\bf 80}, 5489 (1998).
%
%
\bibitem{searcha}{\it Search for Technicolor Particles in W $+$ 2~jet with
    b-tag Channel at CDF}, T. Handa, K. Maeshima, J. Valls, R. Vilar, The CDF
    Collaboration, FERMILAB-CONF-98/016-E, published in Proceedings of
    Workshop on Physics at the First Muon Collider and at the Front End of a
    Muon Collider, Fermi National Accelerator Laboratory, Batavia, IL,
    November 6-9, 1997.
%
%
\bibitem{searchb}{\it Search for a Technicolor $\omega_T$ Particle in Events
with a Photon and a b-quark Jet at CDF}, F. Abe et al., The CDF
Collaboration, Fermilab-PUB-98/321-E, submitted to Physical Review Letters,
October 1998.
%
%
\bibitem{searchc}{\it CDF Searches for New Phenomena.}, D. Toback, The CDF
Collaboration, FERMILAB-CONF-98/183-E, published in Proceedings 12th Les
Rencontres de Physique de la Vallee D'Aosta: Results and Perspectives in
Particle Physics, La Thuile, Italy, March 1-7, 1998;\hfil\break
{\it Searches for Exotic Particles at the Tevatron.}  C. Grosso-Pilcher, The
CDF Collaboration, FERMILAB-CONF-98/306-E. Published Proceedings 29th
International Conference on High Energy Physics (ICHEP 98), Vancouver,
British Columbia, Canada, July 23-29, 1998.
%
%
\bibitem{tcsmrates}K.~Lane, {\it Technicolor Production and Decay Rates in
    the Technicolor Straw Man Model}, hep-ph/9903372, Boston University
    Preprint BUHEP-99-5, March 1999.
%
%
\bibitem{mwa}S.~Mrenna and J.~Womersley, private communication.
%
%
\bibitem{pythia}T.~Sj\"ostrand, Comp.~Phys.~Com.~{\bf 82}, 74 (1994).
%
%
\bibitem{ehlq}E.~Eichten, I.~Hinchliffe, K.~Lane and C.~Quigg,
  Rev.~Mod.~Phys~{\bf 56}, 579 (1984). 
%
%
\bibitem{mwb}S.~Mrenna and J.~Womersley, {\it Can a light technipion
    be discovered at the Tevatron if it decays to two gluons?}, to be
    published in Physics Letters~B, hep-ph/9901202, Fermilab-PUB-99-002
    (1999).
%
%
\bibitem{mn}M.~Narain, private communication.


\end{thebibliography}
\end{document}